\documentclass[preprint2]{aastex}

\usepackage{amsmath}
\usepackage{amssymb}
\usepackage{graphicx,graphics}
\usepackage[english]{babel}
\usepackage{natbib}
\usepackage{color}
\usepackage{subfigure}

\def\PGPU{$\varphi$GPU }
\def\PGRAPE{$\varphi$GRAPE }

\begin{document}

\title{The Link Between Ejected Stars, Hardening and Eccentricity Growth of Super Massive Black Holes in Galactic Nuclei}

\author{Long Wang\altaffilmark{1,2}, 
        Peter Berczik\altaffilmark{3,4,5}, 
        Rainer Spurzem \altaffilmark{3,4,2} and
        M.B.N. Kouwenhoven\altaffilmark{2,1}}
  
\altaffiltext{1}{Department of Astronomy, School of Physics, Peking University, Yiheyuan Lu 5, Haidian Qu, 100871, Beijing, China}      
\altaffiltext{2}{Kavli Institute for Astronomy and Astrophysics, Peking
  University, Yiheyuan Lu 5, Haidian Qu, 100871, Beijing, China}
\altaffiltext{3}{National Astronomical Observatories of China, Chinese Academy of Sciences, 20A Datun Rd., Chaoyang District, 100012, Beijing, China}
\altaffiltext{4}{Astronomisches Rechen-Institut, Zentrum f\"ur Astronomie, University of Heidelberg, M\"onchhofstrasse 12-14, 69120, Heidelberg, Germany}
\altaffiltext{5}{Main Astronomical Observatory, National Academy of Sciences of Ukraine, 27 Akademika Zabolotnoho St., 03680, Kyiv, Ukraine}

\shorttitle{Ejected Stars & Binary Black Holes}
\shortauthors{Wang et al.}

\begin{abstract}
The hierarchical galaxy formation picture suggests that super massive black
holes (MBHs) observed in galactic nuclei today have grown from coalescence of
massive black hole binaries (MBHB) after galaxy merging. Once the components of
a MBHB become gravitationally bound, strong three-body encounters between the
MBHB and stars dominate its evolution in a ``dry'' gas free environment, and
change the MBHB's energy and angular momentum (semi-major axis, eccentricity and
orientation).  Here we present high accuracy direct $N$-body simulations of
spherical and axisymmetric (rotating) galactic nuclei with order $10^6$ stars
and two massive black holes that are initially unbound. We analyze the
properties of the ejected stars due to \textit{slingshot} effects from
three-body encounters with the MBHB in detail. Previous studies have
investigated the eccentricity and energy changes of MBHs using approximate
models or Monte-Carlo three body scatterings. We find general agreement with the
average results of previous semi-analytic models for spherical galactic nuclei,
but our results show a large statistical variation.
Our new results show many more phase space details of how the process works, and
also show the influence of stellar system rotation on the
process. We detect that the angle between the orbital plane of the
  MBHBs and that of the stellar system (when it rotates) influences the 
phase-space properties of the ejected stars. We also find that massive MBHB 
tend to switch stars with counter-rotating orbits into co-rotating orbits during
 their interactions.

\end{abstract}


\section{Introduction}

The current galaxy formation scenario based on the $\Lambda$ cold dark matter
($\Lambda$CDM) cosmology model \citep{Begelman1980,Volonteri2003} suggests that
massive galaxies form from hierarchical merging and accretion of smaller
galaxies. If the merging galaxies have massive black holes (MBHs) in their
centers, the merging processes should also include coalescence of MBHs, since
MBHs are commonly observed at the centers of most galaxies today
\citep{Greene2009}. Thus, to understand the process of MBHs coalescence is
significant for cosmological structure formation, galaxy formation and MBH
formation.

In the case of gas-poor galaxies merging, the two MBHs have three distinct
evolution phases \citep{Begelman1980}. First, the dynamical friction exerted by
the stars forces the two MBHs to sink toward the galactic center. This phase
continues until the binary reaches the \textit{hard binary separation}
\citep{Merritt2001,Yu2002}. Second, a hard MBH binary (MBHB) orbit will continue
to shrink due to \textit{slingshot} ejection of surrounding stars. Third, when
the MBHB reaches a separation at which the loss of orbital energy caused by
gravitational waves (GWs) emission becomes significant, the MBHB shrinks until
the final coalescence.

There is a challenging problem, often called ``Final Parsec Problem'', during
this three phase model in the merger of gas-poor galaxies. In
\textit{quasi-steady spherical} stellar environment, the \textit{slingshot}
efficiency decreases rapidly when the separation of the MBHB becomes much
smaller than the average separation of neighboring stars of the MBHB
\citep{Quinlan1997, Milosavlijevic2003, Berczik2005}. The hardening rate of the
MBHB for large particle numbers, such as in real galactic nuclei, is too low and
the timescale of coalescence of the MBHB will be larger than the Hubble
time. Recent efforts suggest that if the stellar system has some degree of
axisymmetry or triaxiality, this problem may be solved
\citep{Yu2002, Merritt2004, Berczik2006, Preto2011,FPB2012, KPBB2012, khan2013}.
 The simulations by \cite{khan2013} show that even a mildly flattened stellar system with an axis ratio 
of $0.9$ will result in an eight times faster hardening rate of the MBHB,
and that the MBHB evolution is independent of particle number when the
axis ratio is $0.75$. These results agree with the earlier work of
 \cite{Berczik2006}. Also, \cite{Gualandris2012} argue
that in post-merger galaxies the loss-cone is full, hardening does not 
depend on $N$. Therefore, these studies show a solution of the ``Final Parsec Problem''
because the MBHB merges in a relatively short time.
The final solution of this problem requires more work on the details of the
\textit{slingshot} phase.

Detection of GWs is the most important expected observations in the near future
to give direct proof of general relativity. GWs will also provide a new window
to understand the Universe independent of electromagnetic observations. The
coalescence of MBHs is expected to be the strongest GW source to be measured
with future space satellite detectors, such as the Laser Interferometer Space
Antennae (LISA/e-LISA/ALIA, see e.g. \citealp{Gong2011}). This encourages
researchers to concentrate on the final orbital parameters and the merging rate
of MBHs to predict the GW signals to be measured.

It is very important to understand the eccentricity growth of the MBHB during
phase $2$ since it strongly influences the coalescence time \citep{Peters1964}
and the angular momentum evolution. \cite{Q1996} used three-body scattering
experiments to study the properties of slingshot effects. He found the
eccentricity growth is small unless the MBHB already forms with a large
eccentricity. \cite{Preto2011,khan2011} carried out a large number of $N$-body
simulations of \textit{equal-mass} MBHBs in axisymmetric and triaxial stellar
environment after a galaxy merger. They found the initial eccentricities when
MBHs become bound are very high with about $e = 0.95$ on average. This is also
consistent with previous work \citep{Aarseth2003, BPB2008, Berentzen2009, 
Preto2009, LLB2012}. On the other hand, \cite{khan2013} find low eccentricities for some of their elliptical
 galaxy models. The reason for this may be that they model a non-rotating stellar system. 
However, none of the these studies examined the detailed 
scattering processes near the MBHB; it is desirable to study the eccentricity 
evolution in detail in the full $N$-body simulation as a counterpart, and to 
check the validity of \cite{Q1996} three-body Monte Carlo work.

A galaxy merger can result in a rotating stellar
  system. Therefore, simulations of MBHBs in rotating star clusters are
 interesting \citep{Berczik2006,Preto2011,khan2011,KPBB2012}.
Several restricted studies of MBHB evolution in rotating star 
clusters were carried out by \cite{Sesana2011} and \cite{GDS2012}.
They study very high mass ratios of the MBHB ($1/64$) and a
 small stellar system surrounding it (less than the mass of the MBHB). 
 Under these restrictions, \cite{Sesana2011} find that the 
eccentricity evolution of MBHB in a rotating stellar cusp can be 
significantly influenced by the fraction of corotating stars. 
However, larger $N$-body simulations may give different results,
because many interactions occur with stars from unbound regious.
Thus to understand the effect of stellar rotation further, it is necessary to carry 
out a deeper analysis of large $N$-body systems with different mass ratios 
of MBHB.

It is interesting that triple black holes during galaxy mergers can also exist,
if one of the progenitor MBHBs does not coalesce before the next galaxy with
another third MBH falls in. In such cases, very extreme eccentricities of the
inner MBHB have been found ($e\sim 0.99-0.999$, \citealp{Amaro2010})

\cite{Yu2003} studied the dynamical processes of hyper-velocity stars ejected by
the MBHB in the Galactic center and predicted the rate of
ejections. \cite{Lu2010} use analytical arguments and numerical simulations to
study the distribution of hyper-velocity stars ejected by a MBH in the Galactic
center. They found most of these are ejected anti-parallel to the injecting
direction of their progenitors.

In this paper, Section~\ref{sec-method} provides the $N$-body simulations and
the method to select ejected stars from the simulation results. In
Section~\ref{sec:ui} we list the initial conditions. In Section~\ref{sec:result}
we discuss the eccentricity growth of MBHBs, angular momentum properties of MBHBs
and ejected stars. Finally, Section~\ref{sec:dis} contains our results and
conclusions.


\section{Methods} \label{sec-method}

In this work, we use the direct $N$-body code called \PGPU ~\citep{BNZ2011} 
to do all the simulations. The full set of runs and the parameters are 
described in more detail in our forthcoming publication ~\citep{BSBN2013}.

The code is a direct $N$-body simulation package, with a high order
Hermite integration scheme and individual block time steps (the code supports
time integration of particle orbits with 4$^{\rm th}$, 6$^{\rm th}$ and even
8$^{\rm th}$ order schemes). A direct $N$-body code evaluates in principle
all pairwise forces between the gravitating particles, and its computational
complexity scales asymptotically with $N^2$; however, it is {\em not} to be 
confused with a simple brute force shared time step code, due to the block time
steps. We refer more interested readers to a general discussion about $N$-body codes
and their implementation in \cite{spurzem2011a,spurzem2011b}.

The \PGPU code is fully parallelized using the MPI library, and for each MPI process
GPU accelerator hardware is used to compute gravitational forces between
particles. It is based on an earlier {\bf \tt C}
version\footnote{\tt ftp://mao.kiev.ua/pub/users/berczik/phi-GRAPE/} for
{\tt GRAPE6a} clusters ~\citep{FMK2005}. The new code is written from scratch in
{\bf \tt C++} and based on earlier CPU serial $N$-body code (YEBISU;
\citealp{NM2008} ).  The MPI parallelization was done in the same ``j'' particle
parallelization mode as in the earlier \PGRAPE code \citep{HGM2007}.

The present version of the \PGPU\footnote{\tt ftp://mao.kiev.ua/pub/berczik/phi-GPU/}
code uses native GPU support and direct access
to the GPU's with only the NVIDIA native CUDA library. Multi GPU support is
achieved through MPI parallelization. More details and also the \PGPU public
version are presented in \cite{BNZ2011, SBZ2012, Berczik2013}.

The present code is well tested and already used to obtain important
results in our earlier large scale few million body simulation
~\citep{KBB2012}.

\begin{figure*}[htbp]
  \begin{center}
      \includegraphics[angle=0,width=1.5\columnwidth]{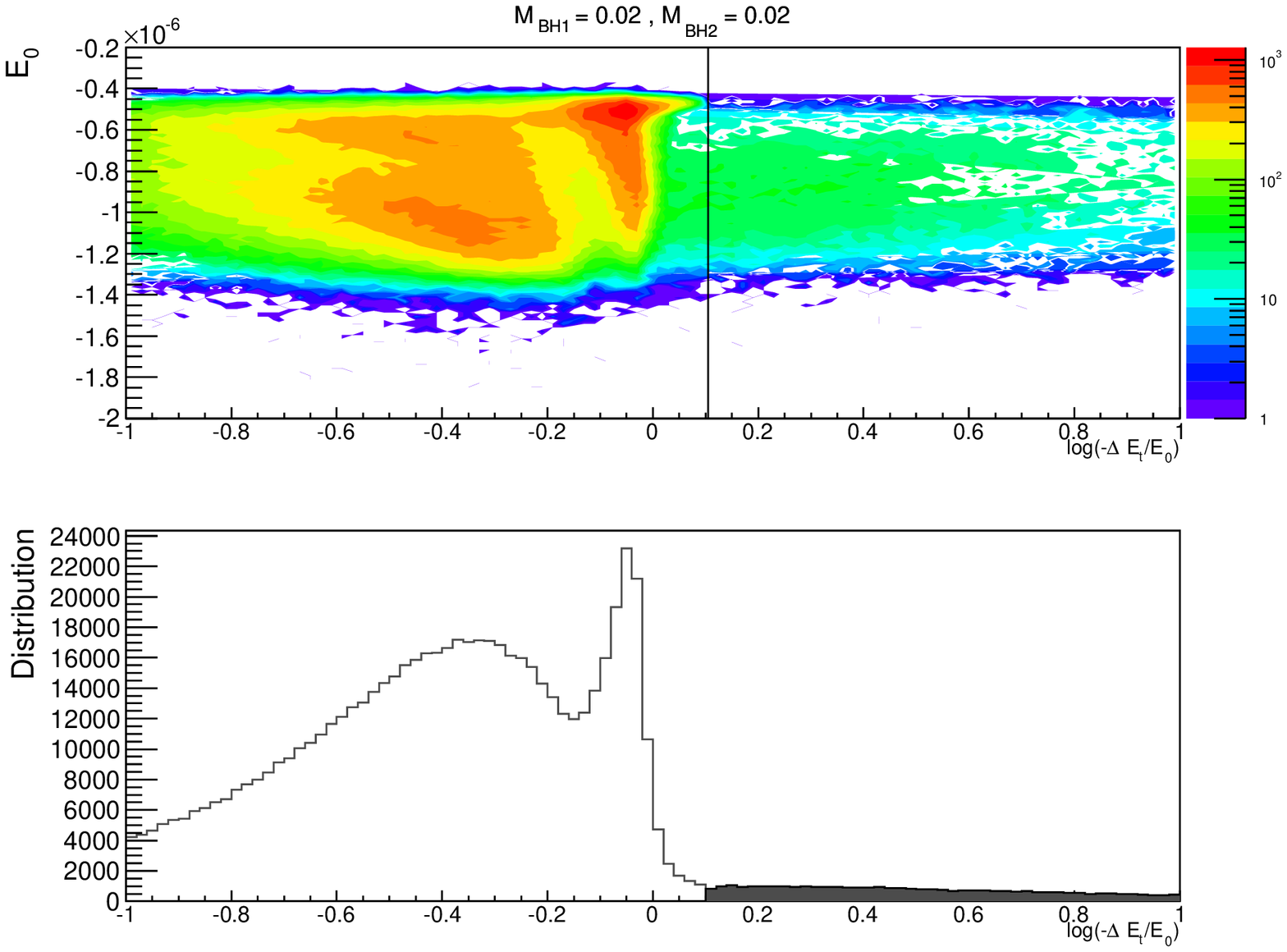} \\
      (a)\\
      \includegraphics[angle=0,width=1.5\columnwidth]{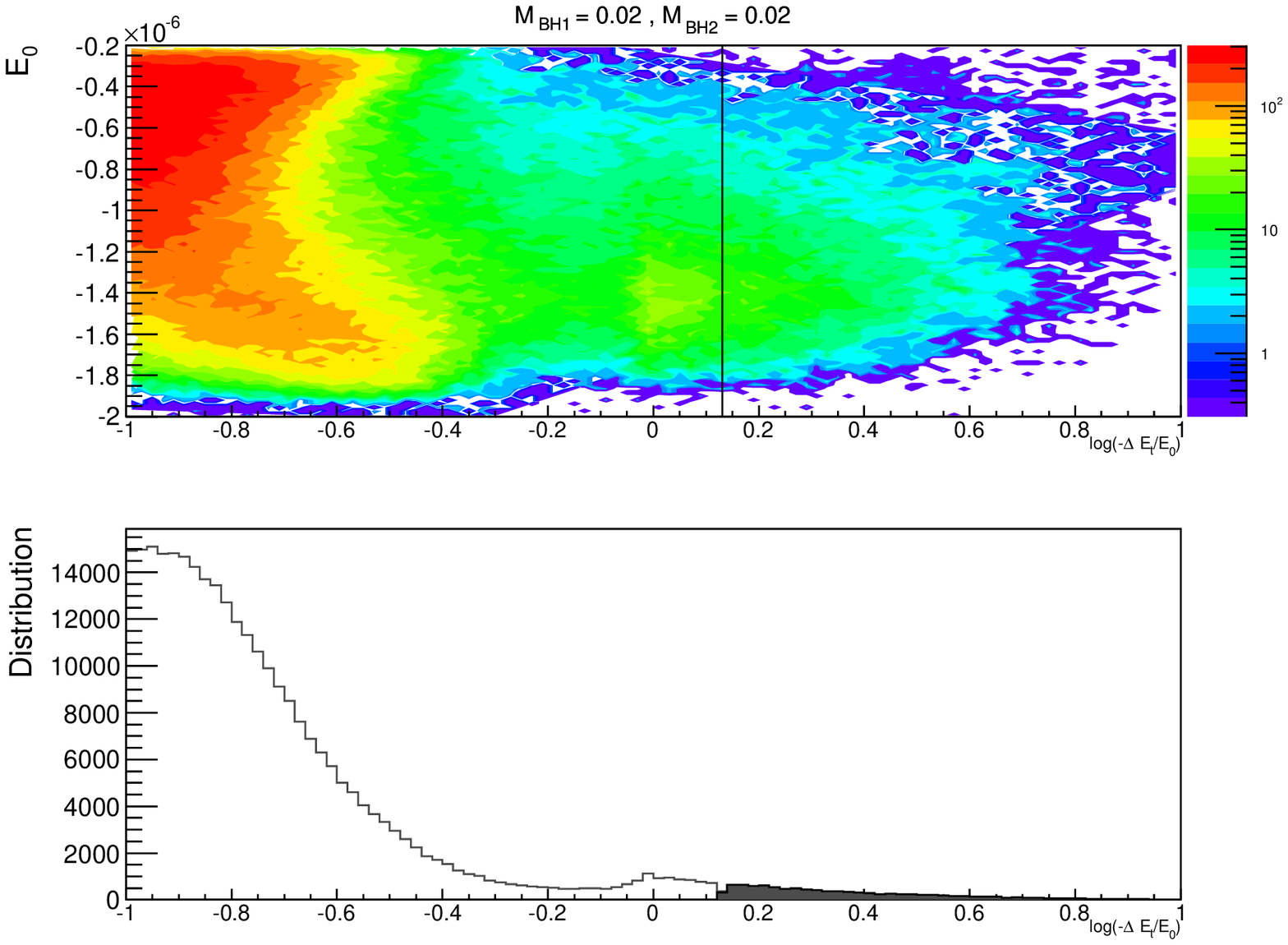}\\
      (b)\\
  \end{center}
  \caption{The total energy change of individual stars for model $2020$ (a) and
    $2020$\textit{-nonrot} (b). the upper panels in (a) and (b) are two-dimension histograms of $E_0$
    vs. $\log(-\Delta E_t/E_0)$. The bottom panels in (a) and (b) show the distributions of $\log(-\Delta
    E_t/E_0)$. The black line is threshold determined by the critical value $\delta_e$
    which separates the two number density populations
    (Section~\ref{sec-method}).  The stars in right regions are selected as
    ejected star candidates.}
  \label{fig:dEEs}
\end{figure*}

Our analysis is based on the ejected stars from the \textit{slingshot} effect of
MBHBs. The method used to select ejected stars depends on the individual total
energy change from the beginning to the end of the simulations ($\Delta E_t$)
because ejected stars usually gain a lot of energy during the interaction and
this energy gain should be much larger than energy fluctuations caused by
perturbations from other stars. The next step is to check the two dimensional
histogram of initial energy of each star $E_0$ vs. $\log(-\Delta E_t/E_0)$ for
each model, and to determine a critial value $\delta_e$ to select ejected star
candidates, which satisfy $\log(-\Delta E_t/E_0)> \delta_e$, where $\delta_e$ is
obtained by the number density gap shown in Figure~\ref{fig:dEEs}. For rotating
and non-rotating models, these energy features are different, but we still can
select a $\delta_e$, which works well. This procedure also functions as an
operational definition of the term ejected star, which here just means it leaves
after a strong encounter with the MBHB with significantly higher energy than before.
For the purpose of our paper it is not important whether the ejected star has
enough energy to become unbound from the MBHB, from the central stellar cluster
or even from the entire galaxy.

The next step is to check the energy evolution of each ejected star candidate
and the MBHB. The final samples of stars are chosen from each candidate energy
change $\Delta E_i > M_{s,i}$ during its ejection time $t_e$, where $M_{s,i}$ is
the mass of a star in $N$-body units and $i$ indicates the index of ejected
stars.  In our simulation, $M_{s,i}$ is the same order of magnitude of the
individual star's energy in $N$-body units. Thus it can be used to select the
events with obvious energy jump.

With the ejected stars samples, we calculate the angular momentum $L_i$ of each
ES and $L_b$ of the MBHB at $t_e$. We compare the angular momentum before and
after ejection of each star.

We carry out all data reduction and analysis using the open source software {\textit{ROOT}}.

\section{Units and Initial Conditions} \label{sec:ui}

We scale the numerical units of our initial models applying the standard $N$-
body normalization \citep{AHW1974} by setting both the gravitational constant
$G$ and the total mass of the stellar system to unity. The total energy of the
system is scaled to $E = - {1/4}$. Our simulations are purely gravitational and
thus scale-free, but for convenience we define one example of the scaling in
physical units below. The initial conditions of our $N$-body simulations
presented here are based on the ones used in \cite{Berczik2006} and
\cite{BPBMS2009}. The initial stellar galactic nucleus follows a distribution
function for a rotating King model \cite[see, e.g.,][and references
  therein]{ES1999}. It provides rigid rotation inside the half-mass radius, and
quickly decreasing differential rotation outwards. After the MBHs settle into
the galactic center the density and velocity dispersion adjust to the MBH's
gravity inside its influence radius.  The concentration and rotation parameters
are set to $W_0 = 6$ and to $\omega_0=1.8$, respectively, in all rotating
models. We also simulate the non-rotating King models ($\omega_0=0$) for
comparison. The total angular momentum vector of the stellar nucleus are aligned
with the $z$-axis of our coordinate frame. We place the two MBHBs in the $z=0$
mid-plane with initial coordinate components $x_{1,2}=0$ and $y_{1,2}=\pm 0.3$,
where the subscripts denote the two black hole particles.

The full set of models (with different field particle numbers and set of MBH's
initial velocities) and the MBH's orbital evolution are presented elsewhere
\citep{BSBN2013}. Here we analyze only the subset of our runs (which include
seven models) with a fixed stellar particle number $N = 10^6$. The total
integration time for these models was $150$ $N$-body time units. The differences
are the MBH masses, which we are given in Table~\ref{tab:mod}.  The
non-rotating models are indicated by the suffix \textit{-nonrot} hereafter. Our
work will focus on the rotating models. Thus non-rotating models will only be
shown in some parts. The initial $x$-velocity of the MBH's in these simulations
has been chosen to be $v_{x;1,2}=\pm V_{\mathrm{circ}}$, where
$V_{\mathrm{circ}}$ is the circular velocity within the stellar background
model. With our choice of initial values, the circular velocity in $N$-body
units is $V_{\mathrm{circ}}=0.7$ at the initial distance of the MBH's from the
center.

Our models are scale-free and can be applied to a range of real astrophysical
systems. Here we give an example, for the case of the MBH mass 0.01 in $N$-body units i
(see Table~\ref{tab:mod}); if the black hole mass
is e.g. $10^7 M_\odot$, and the black hole separation $y_{1,2}=\pm 0.3$ is,
e.g., $\pm 300$ pc, then one $N$-body time unit is about $15$ Myr and one
$N$-body velocity unit is about $65.6 $ km/s.

\begin{table*}[htbp]
  \begin{center}
    \caption{Full set of our model runs for both rotating and non-rotating King models
    (we use $W_0 =6$ and $\omega_0=1.8$ for the rotating and $\omega_0 = 0$ for the
 non-rotating models).
 $m_1$, $m_2$ and $\mu = m_1 m_2 / M$, with $M = m_1 + m_2$ denote the masses of
the primary and secondary MBH and their reduced mass; all masses here are in units
of $10^{-2}$, where the total cluster mass is unity. \medskip }
    \begin{tabular}{l|ccccccc}
      \tableline
      Model & 0110 & 0210 & 0510 & 1010 & 2020 & 4020 & 4040 \\
      \tableline
      $m_2/m_1$ & 0.10/1 & 0.20/1 & 0.50/1 & 1.00/1 & 2.00/2 & 4.00/2 & 4.00/4 \\
      $\mu$ & 0.0909 & 0.1667 & 0.3333 & 0.5000 & 0.6667 & 1.3333 & 2.0000 \\
      \tableline
      \end{tabular}
  \label{tab:mod}
  \end{center}
\end{table*}

\section{Results} \label{sec:result}

\subsection{Coordinate System and Angles} \label{sec:csa}

In our simulations, we have initially defined rectangular, Cartesian coordinates with
$x$, $y$ and $z$ axes (Section~\ref{sec:ui}). Here, we define three angles:
$\alpha$, $\delta$ and $\theta$. When we use the spherical coordinate system
instead, the radius $r$ denotes the distance to the origin and two angles $\alpha$ and $\delta$
define the direction of a vector. 
The transformation from $(x,y,z)$ to $(r, \alpha, \delta)$ can be described as:
\begin{equation}
  \begin{split}
  x &= r \cos \delta \cos \alpha \\
  y &= r \cos \delta \sin \alpha \\
  z &= r \sin \delta \\
  \end{split}
\end{equation}
We will later use these spherical coordinates to define the angular momentum vectors
of a star $L_s$ and of the MBHB $L_b$.
$\theta$ is the angle between $L_b$ and $L_s$ defined through:
\begin{equation}
  \theta = \arccos \frac{\overrightarrow{L_b} \cdot \overrightarrow{L_s}}{|\overrightarrow{L_b}| |\overrightarrow{L_s}|}
\end{equation}
 Hereafter we use suffix ``b'' to denote MBHBs and ``s'' to denote ejected
 stars.

For each angle of individual ejected stars or MBHBs, we also have two values:
the one before ejection time $t_i$ (hereafter denoted with suffix ``BE'') and the
one after ejection time $t_o$ (hereafter denoted with suffix ``AE''). Due to our
simulation output time resolution, the interval time between $t_i$ and $t_o$ is
one $N$-body time unit.

\subsection{Ejected Stars Sample Selection}

Using the method discussed in Section~\ref{sec-method}, we successfully find
most ejected stars for the massive MBHB models. The numbers of ejected stars and
$\delta_e$ (defined in Section~\ref{sec-method}) for all rotating models are listed in
Table~\ref{tab:table-eject}.

\begin{table*}[htbp]
  \begin{center}
    \caption{The number of ejected stars detected in each rotating model. \medskip}
    \begin{tabular}{l|ccccccc}  
      \tableline
      Model           & 0110 & 0210  & 0510   & 1010   & 2020    & 4020   & 4040 \\
      \tableline
      $\delta_e$    & 0.25  & 0.25   & 0.24    & 0.20    & 0.24     & 0.19   & 0.08 \\
      $N_{ESs}$       & 863   & 3457  & 10203 & 16656 & 40288  & 57596  & 83367 \\    
      $N_{ESs}/N_{tot}(\%)$   & $0.086$ & $0.34$ & $1.02$ & $1.67$ & $4.03$ & $5.76$ & $8.34$ \\
      \tableline
    \end{tabular}
  \label{tab:table-eject}
  \end{center}
\end{table*}

\begin{figure*}[htbp]
  \begin{center}
      \includegraphics[angle=0,width=1.5\columnwidth]{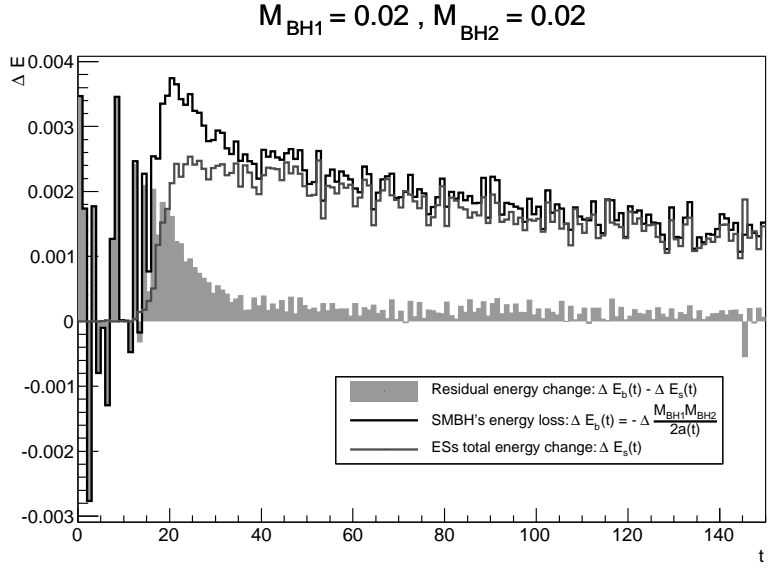}\\
      \includegraphics[angle=0,width=1.5\columnwidth]{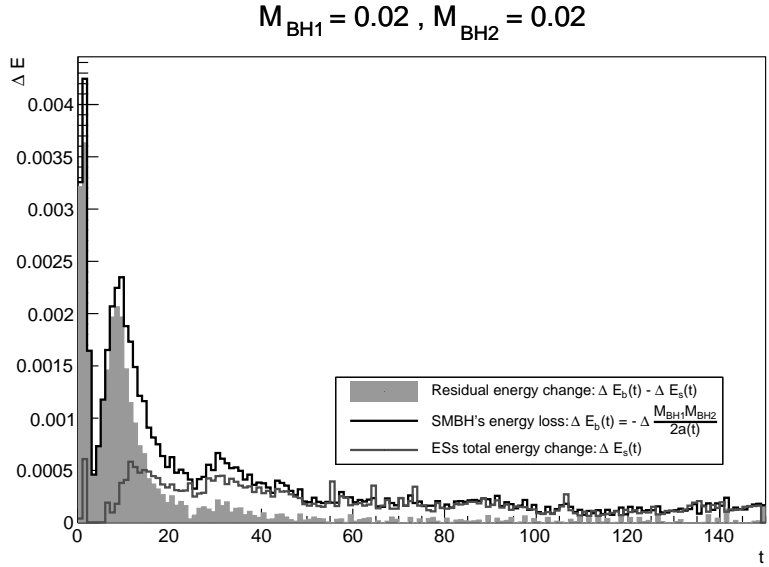}\\
  \end{center}
  \caption{The histogram of energy change of MBHBs, ejected stars and residual energy with
    respect to time, for model $2020$ and $2020$\textit{-nonrot} (see Table~\ref{tab:mod}). $\Delta E_s(t)$ is the integration of energy
    change of ejected stars ejecting bewteen time $t$ and $t+1$ ($N$-body unit).}
  \label{fig:dEvsT}
\end{figure*}

To ensure our samples are convincing, we calculate the integrated energy change
of all ejected stars during their ejections ($\Delta E_{s}(t)$) and compare it to the MBHBs'
binding energy loss ($\Delta E_{b}(t)$) during each time unit
(Figure~\ref{fig:dEvsT}). If the residual energy change $\Delta E_{r}(t) =
\Delta E_{b}(t) - \Delta E_{s}(t)$ is zero, the ejected star sample
is complete, its deviation from zero gives information about how many
ejected stars we may have missed, since there is no other significant mechanism in our
simulations due to which the MBHB can lose energy. 
Figure~\ref{fig:dEvsT} shows that $\Delta E_{r}(t)$ in both the
rotating model $2020$ and the non-rotating model $2020$\textit{-nonrot} are
almost zero after the binary formation ($t>40$). This result holds for all
models with large black hole masses. This indicates our method to select
ejected stars is reliable. The only exceptions are the two models $0110$ and
$0210$ with low masses of MBHBs (one is shown in Figure~\ref{fig:dEvsT}).

There are two possibilities that may cause incomplete samples for low mass MBHBs. One
is that low mass MBHBs probably generate more ejected stars with low $\Delta
E_i$, which cannot be distinguished from energy fluctuations caused by other mechanisms, so we
cannot select this part of the ejected stars. Another reason is that the low mass MBHB's
become gravitational bound at a later time
than the massive ones -- slingshot effects dominate the
energy loss of MBHBs only after the binary formation.

\subsection{Eccentricity Growth Rate of MBHBs}

The specific angular momentum $J$ of a MBHB can be described by
\begin{equation}\label{eq:l}
  J = \frac{L}{\mu} = \sqrt{ GMa (1-e^2)}
\end{equation}
 where $L$ is the standard angular momentum, $\mu$ is reduced mass, $a$ is
semi-major axis, $M = m_1 + m_2$ is the total mass of the binary components, $e$ is
the eccentricity and $G$ is gravitational constant. The hardening process of a MBHB provides energy to the
ejected stars, and thus increases its binding energy and reduces its semi-major axis $a$;
As a result of Eq.~\ref{eq:l} the angular momentum of the MBHB will also be reduced,
even if $e$ remains constant (which is generally not the case; see below).
Any eccentricity growth will lead to additional decrease of $L$.
In phase $2$, the ejected stars 
dominate the energy and angular momentum evolution of MBHB. Thus to understand
the properties of ejected stars it will help to know how they carry away the
energy and the angular momentum from the MBHB.

\cite{Q1996} defined an eccentricity growth rate $K$ as
\begin{equation}
  \label{eq:ke}
  K = \frac{\Delta e }{\Delta \ln (1/a)}
\end{equation}
and then derived $K_1$, a numerical expression of $K$ with the assumption that all stars have an identical velocity
$v$. Note that, for the special case of constant specific angular momentum
($\Delta J = 0$), we have
\begin{equation}
  \label{eq:ke1}
  K = \frac{\Delta e }{\Delta \ln (1/a)} = - \frac{1-e^2}{2e} \ .
\end{equation}
Quinlan also carried out Monte-Carlo models (three-body scattering experiments of
single stars with the MBHB), the results of which we can compare with our data. 
In our work, we
also calculate $K$ by using Equation~\ref{eq:ke} directly from the MBHBs measured
changes $\Delta e$ and $\Delta \ln (1/a)$. To compute the differences we use
the eccentricity $e$ and semi-major axis $a$ of the MBHB averaged over time spans
of one $N$-body time unit (for good statistic).

\begin{figure*}[htbp]
  \begin{center}
    \includegraphics[angle=0,width=2.0\columnwidth]{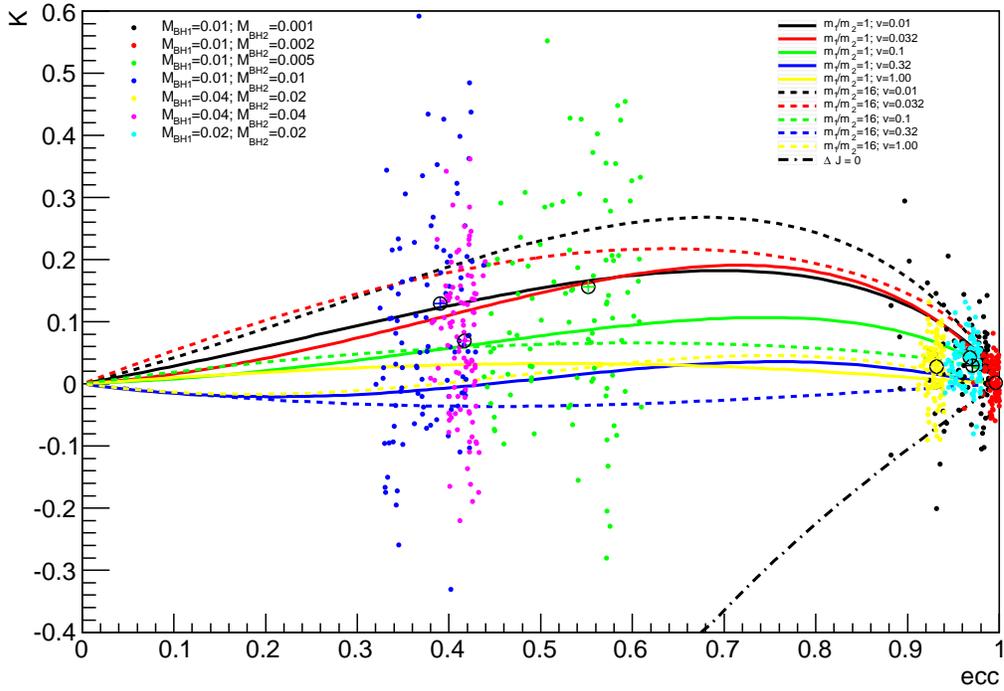}
  \end{center}
  \caption{The eccentricity growth rate $K$ vs. eccentricity $e$ for all rotating
    models. Points are calculated results using Equation~\ref{eq:ke}. Circles
    with crosses are average $K$ between $t=81$ $N$-body units to the end of
    simulation. Solid and dashed curves are fitted function for $K_1$ from
    \cite{Q1996}. Here $v$ represents $v/V_{bin}$ in \cite{Q1996}.}
  \label{fig:ke}
\end{figure*}

Figure~\ref{fig:ke} shows the comparison between our result with \cite{Q1996}'s.
The values are not monotonous in time, rather there is a stochastic variation of
$K$ (which creates positive and negative $\Delta e$ in individual encounters)
due to the three-body encounters. The average growth rate of $e$ (see circles
with crosses in Figure~\ref{fig:ke}) is positive and agrees fairly well with
previous semi-analytic work. The dash-dotted line shows the predicted growth rate
when $J$ of MBHB is conserved. If $J$ of MBHB decreases as the MBHB become harding,
the $K$ should be located above this line. Our results indicate that during the hardening
of the MBHB, the MBHBs with lower $e$ always lose $J$ and the MBHBs with
higher $e$ in most cases also lose $J$.

\subsection{Angular Momentum exchange between stars and MBHBs} \label{sec:am}

\subsubsection{Angular momentum evolution}

\begin{figure*}[htbp]
  \centering
    \begin{tabular}{c c}  
      \includegraphics[angle=0,width=1.0\columnwidth]{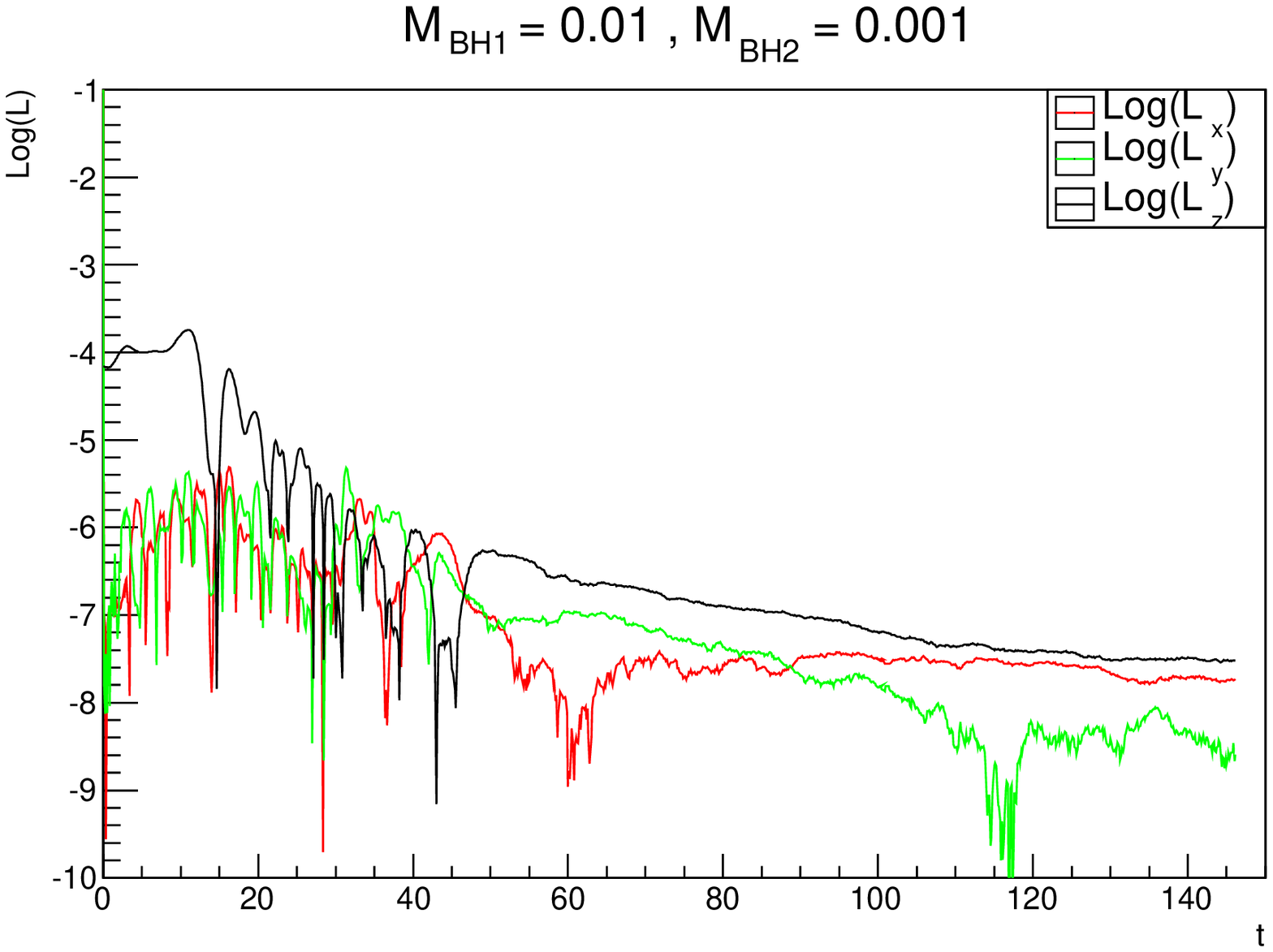}&
      \includegraphics[angle=0,width=1.0\columnwidth]{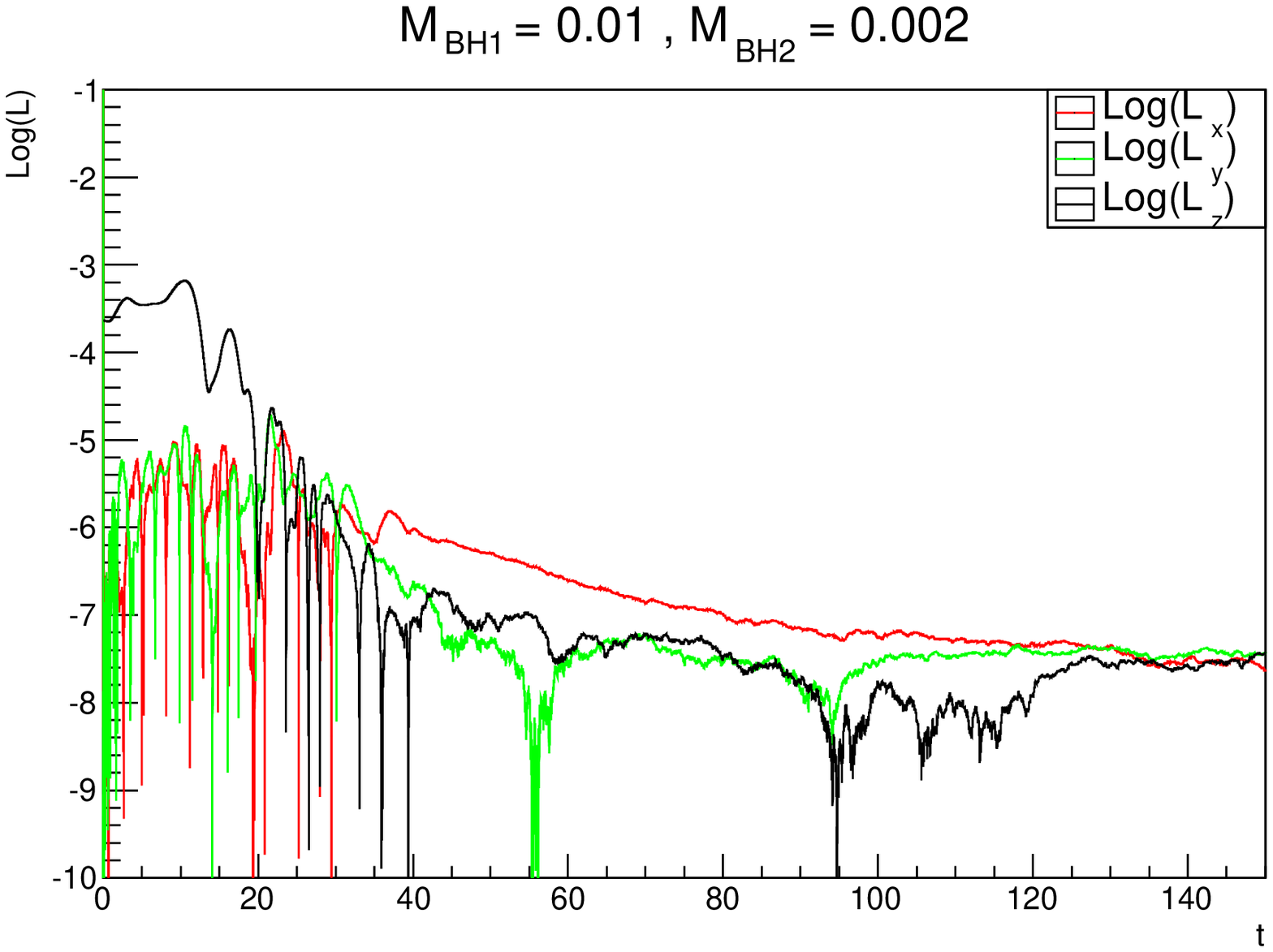}\\
      \includegraphics[angle=0,width=1.0\columnwidth]{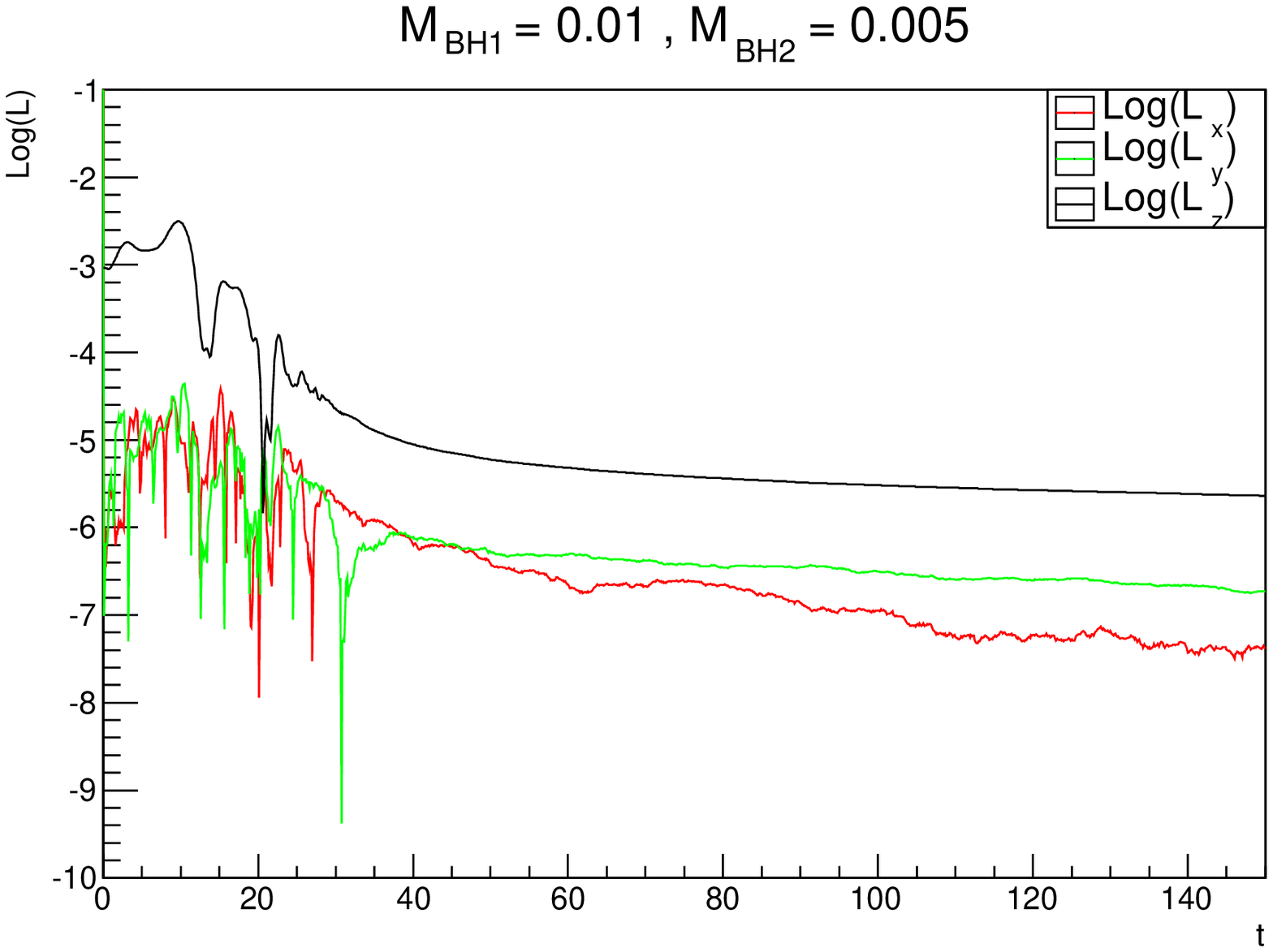}&
      \includegraphics[angle=0,width=1.0\columnwidth]{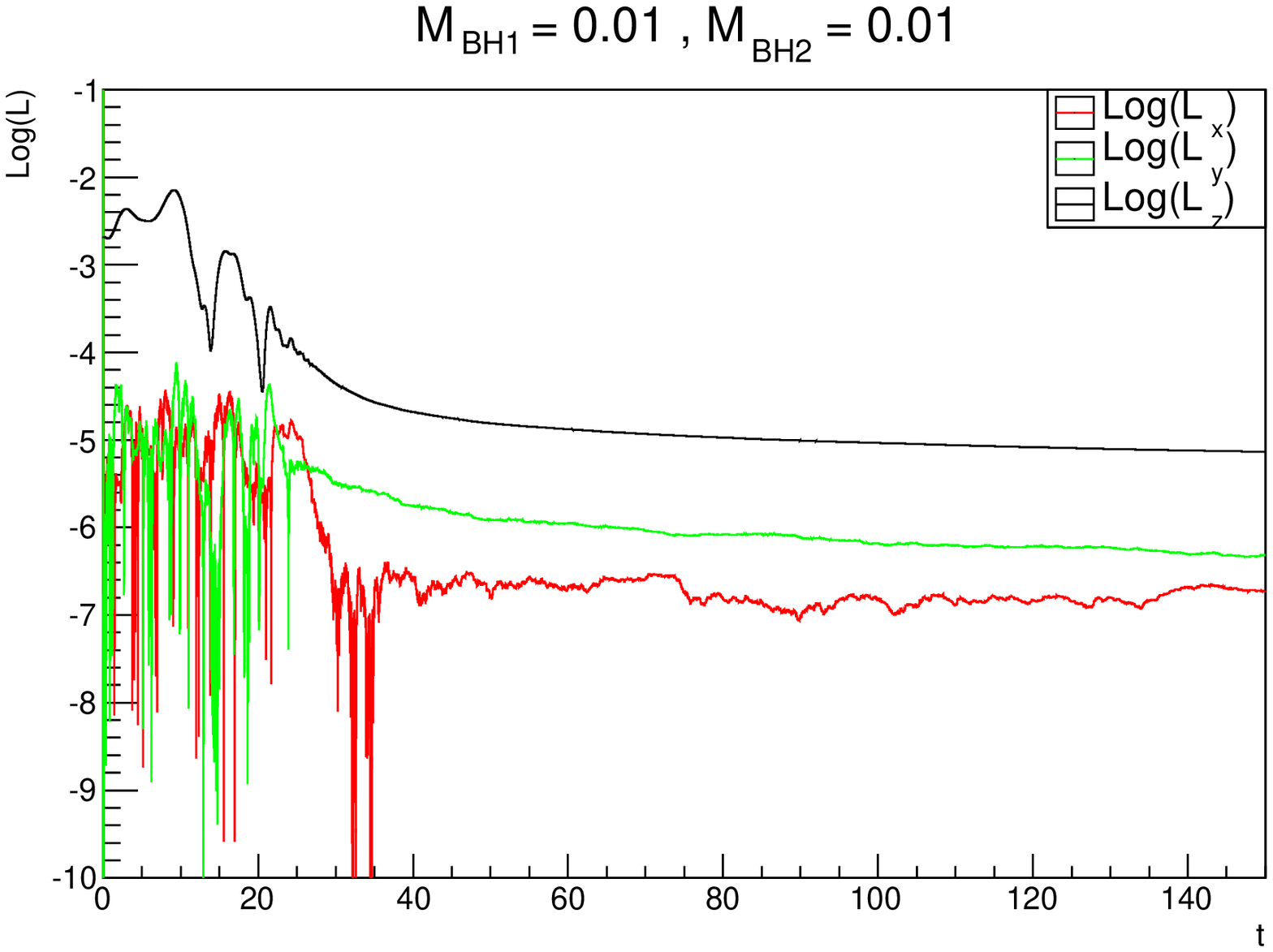}\\
      \includegraphics[angle=0,width=1.0\columnwidth]{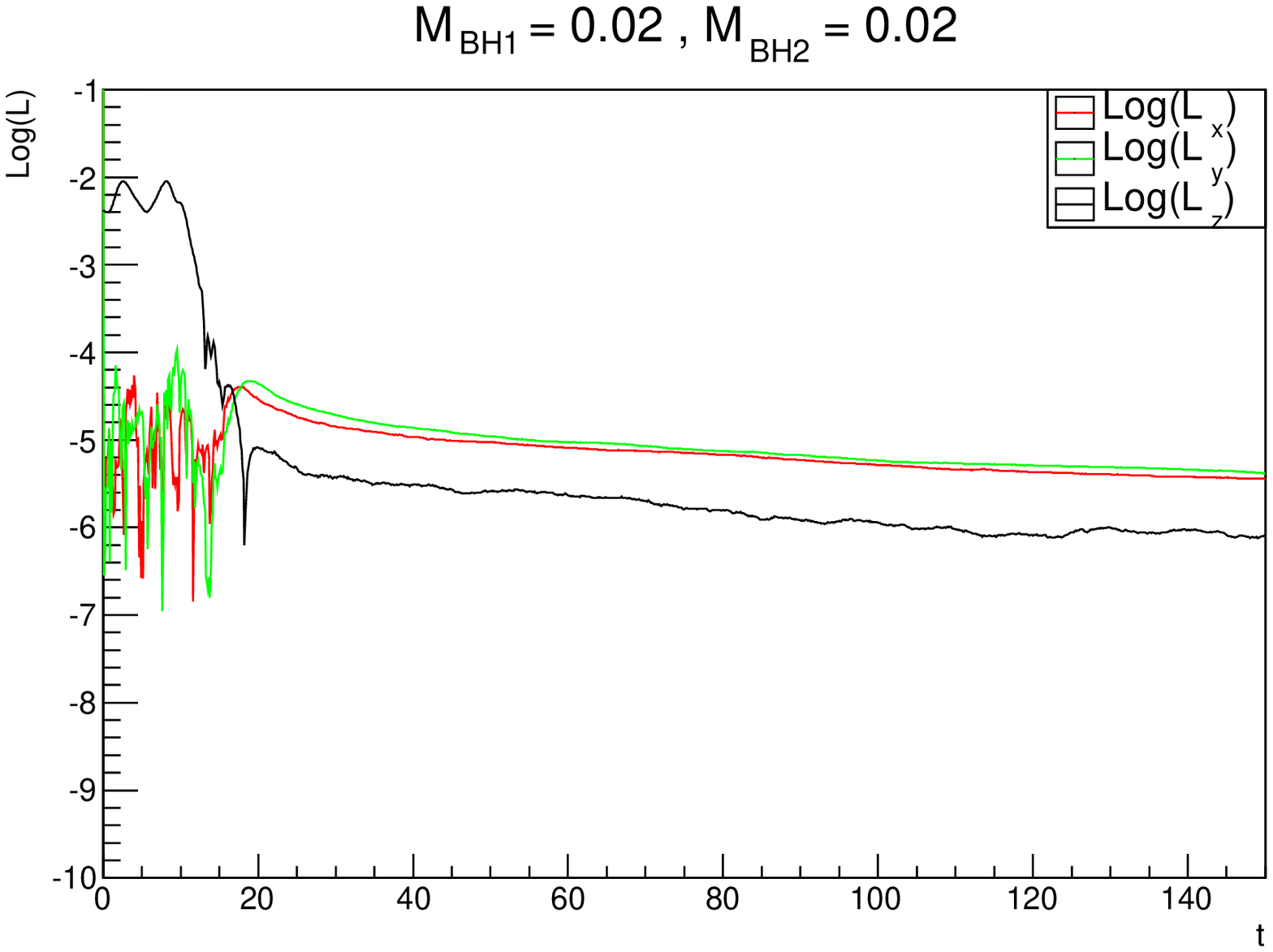}&
      \includegraphics[angle=0,width=1.0\columnwidth]{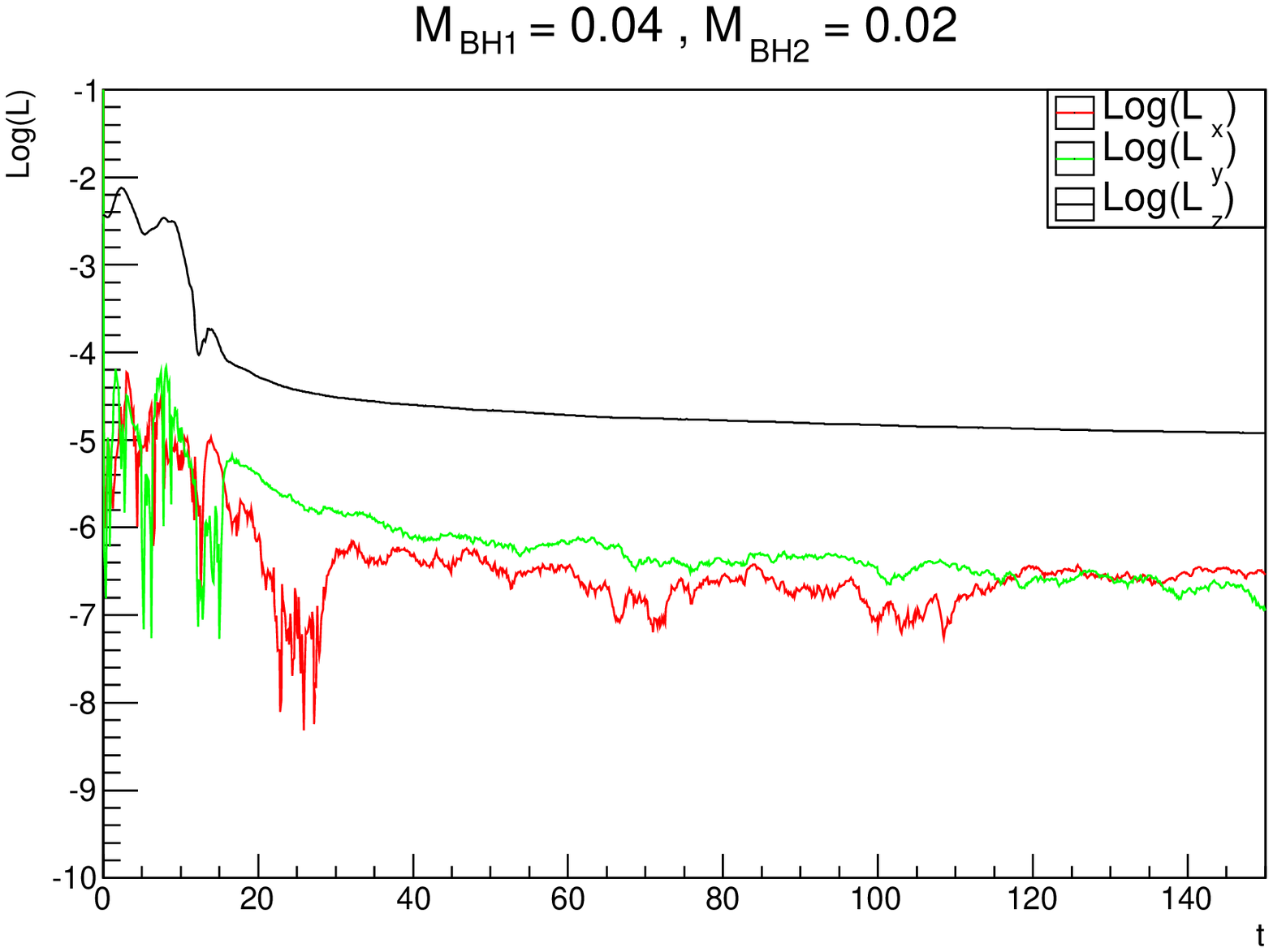}\\
    \end{tabular}
  \caption{The evolution of each components $L_x$ (red), $L_y$ (green) and $L_z$
    (black) of the MBHBs angular momentum(Section~\ref{sec:csa}). The last two panels are
    non-rotation models for comparison. Due to the inclination between the MBHB's
    orbit and stellar system's rotational symmetry plane, models $1001$, $1002$ and $2020$
    are classified as I-models and all others are classified as P-models}
  \label{fig:bh_xyz}
\end{figure*}

Figure~\ref{fig:bh_xyz} shows how the three components of the
MBHBs' angular momentum decrease over time. The $L_z$ of
MBHBs in models $0110$, $0210$ and $2020$ have the same or smaller magnitude of
$L_x$ and $L_y$. Thus their MBHBs have an orbital plane that is tilted with respect to the
$x-y$ plane, which is the symmetry plane of the rotating stellar cluster.
The inclination angle between these two planes is far from zero,
hereafter we call these models I-Models. In contrast, the MBHBs in all the other
models have an orbital plane almost parallel to the $x-y$ plane (hereafter we
call these P-Models).

The angular momentum of individual ejected stars $L_{r,AE}$ as compared to that
before ejection $L_{r,BE}$ in all models shows an increasing trend.
Figure~\ref{fig:L_io} provides evidence for this trend. If one of the ejected stars
gains $L_r$ during its ejection, it is located in the top-left region of the
density map, and vice versa. We see that there is both gain and loss of $L_r$ of
ejected stars. But the total $L_r$ gain of the ejected stars is larger than the loss
since the density peaks are located in the top-left region for all models. This
indicates that the ejected stars will carry away net angular momentum from the MBHB.
As an additional effect, they also carry away and redistribute some of the angular
momentum of the stellar system.

Two more effects are shown in Figure~\ref{fig:L_io}. One is that,
independent of the mass of the MBHBs, the stellar angular momentum after
the encounter $L_{r,AE}$ is approximately constant (it is actually a distribution
where the highest level contour lines are nearly flat, parallel to the horizontal axis,
which is the angular momentum before the encounter $L_{r,BE}$). This means
that stars of any incoming angular momentum get a typical angular momentum
after the encounter which is independent of its initial value and is only determined
by the properties of the MBHB. The value of such post-encounter angular momentum becomes smaller
for larger MBHB mass. In case of the non-rotating stellar system the effect is
not visible in the plot, the angular momentum after the encounter scatters
in a more symmetric distribution around the line of equality with the initial
angular momentum.

\subsubsection{Distribution of angular momentum direction}

The direction of incoming and ejecting orbits of ejected stars viewed in
rectangular coordinate system is influenced by rotational planes of both the whole stellar system
and the MBHB. The distribution of $\alpha_s$ and $\delta_s$ (see
Section~\ref{sec:csa}) is similar before and after ejection
(Figure~\ref{fig:ad}).

For I-Models, the MBHBs have a stable $\alpha_b$ and $\delta_b$ (e.g., $\alpha_b
\approx 3.8$ and $\cos (\delta_b) \approx 0.15$ in model $2020$)
(Figure~\ref{fig:ad}), i.e. the direction of the angular momentum of
the MBHB binary (or its orbital plane) does not change much during all the
stellar encounters.
The $\alpha_s$ and $\delta_s$ distribution concentrates near the same
angles or ($\pm \pi$) as the MBHB, both before and after ejection. 
This indicates that the MBHB interacts preferentially with stars
having the same orbital plane. The rotational direction of the stars
may be the same or opposite rotational direction as MBHBs' (co- or 
counterrotating) and
$\cos (\delta_s)$ also has a concentration towards $0$. Thus ejected
stars are oriented with their orbital plane to the one of the MBHB,
which is
perpendicular to the stellar system rotational symmetry
plane (the $x-y$ plane).

For P-Models, the MBHBs orbital plane is close to the $x-y$ plane,
aligned with the stellar system's rotational symmetry plane and $\alpha_b$ cannot be
defined well. Therefore there is an extended distribution of $\alpha_b$ and we expect
to see no special trend of $\alpha_s$ related to $\alpha_b$ (like model $4020$
in Figure~\ref{fig:ad}). The $\cos (\delta_s)$ distribution also has strong concentration
close to $0$ and less concentration on $\cos (\delta_b)$. The orbits 
of the ejected stars, which prefer 
$\cos (\delta_s) \approx 0$, are now not correlated 
with the MBHB's orbital plane, but rather with the stellar system's rotational
symmetry plane. Therefore we see in this case an effect of the rotation of
the stellar system, which dominate the orbit direction of ejected
stars in P-Models and overrides the MBHBs rotational effect we see in the I-Models.

In Figure~\ref{fig:ad_io} 
we use the angles $\alpha_s$ and $\delta_s$ to illustrate the 
relation between the stellar orbit before (BE) and after (AE) the encounters.
Particularly interesting are the
concentrations near the lines $\alpha_{s,AE}-\alpha_{s,BE}=\pm \pi$, which
indicates that initially counter-rotating stars (with respect to the MBHB orbital
plane) become co-rotating after the encounter and ejection. But
the $\alpha_{s, BE}$ and $\alpha_{s, AE}$ in both I-Models and P-Models show also
concentrations near the line without change, which means that many ejected stars
preserve their rotational direction during the encounter, independent of that of the MBHB.
We can also
see that $\alpha_{s}$ concentrates near $\alpha_{b}$ with small change before and
after ejection in Model $2020$. The $\delta_{s}$ have a wider change than the
$\alpha_s$ and concentrate near $\delta_{s} = \pi/2 $ with a small change.

The information about the relation between incoming and ejected stellar
orbital plane, related to the one of the MBHB, can be more easily analyzed
by looking at a single angle $\theta$, which is the inclination angle
between ejected stars and MBHB (Figure~\ref{fig:theta_io}).
For I-Models, the distribution of
$\cos \theta$ is flat (model $1001$, $1002$) with a little increase around $\cos \theta = \pm 1$
(model $2020$). This indicates that ejected stars show a preference of co- and
counter-rotation with respect to the MBHBs' rotation, which is consistent with the results
discussed above.

For P-Models, $\cos (\theta_{BE})$ and $\cos (\theta_{AE})$ show clear
concentrations near $\pm 1$ and $0$ (slight bias for $\cos (\theta_{AE})$ in model
$4040$) . It means that the ejected stars tend to have a incident and ejecting
orbits parallel or perpendicular to the MBHB's rotational orbit in the
P-Models, which is consistent with Figure~\ref{fig:ad}. 

There is also a slight trend that ejected stars prefer to co-rotate with MBHBs
since the fraction of positive $\cos (\theta)$ is larger than the negative fraction.

The distribution $\cos (\theta)$ in models $4040$ and $4020$ also shows a
significant difference between before and after ejection
(Figure~\ref{fig:theta_io})
. In these two cases, MBHBs tend to switch orbits of
incident stars counter-rotating with MBHBs to co-rotating ones.

For non-rotating models, both $\cos (\theta_{BE})$ and $\cos (\theta_{AE})$ show
no trend of concentration near the $-1$ and $0$, but a strong concentration on
$1$. This indicates that for non-rotating models, ejected stars prefer to have 
incident and ejecting orbits co-rotating with MBH's rotational orbit and also
confirms that the concentration near $0$ is the effect of stellar system
rotation. There is also the trend of switching orbits of incident stars
counter-rotating with MBHBs to co-rotating orbits for rotating models.


\section{Conclusions}\label{sec:dis}

High accuracy direct $N$-body models of spherical and axisymmetric (rotating)
star clusters in galactic nuclei have been presented here, which consist of
one million stars and two massive black holes (MBH), which are initially
unbound. We study the evolution of a massive black hole binary (MBHB) forming during the
evolution and its detailed interactions (superelastic scatterings) with single stars.
The two MBHs have three evolutionary phases: the dynamical friction phase, the three-body
encounter phase and a final gravitational wave (GW) radiation phase. The MBHs will sink toward the galactic
center, form a binary whose orbit shrinks through superelastic three-body encounters 
until they final coalesce under strong emission of gravitational waves. 

Some authors have reported a ``Final Parsec Problem'' for this three phase MBHB
merging scienario purely based on stellar dynamical processes. The timescale of MBHB merging 
would be too long compared to the evolutionary time scale of galactic nuclei and galaxy
mergers, which are the origin of MBHB. The MBHB would stall at a separation of about
a parsec with an empty loss cone, no further hardening (orbit shrinking) occurs and
relativistic energy losses are yet too small \citep{Begelman1980}.

Currently it seems that the ``Final Parsec Problem'' only occurs under unphysical idealized 
conditions, such as a strictly spherical stellar system. Under more general conditions, 
such as some degree of rotation or triaxiality (bars or tidal fields) or the presence 
of gas there is no problem to bring an MBHB in few Gyrs to complete relativistic 
coalescence \citep{Berczik2005, Berczik2006, Preto2011, khan2011, KPBB2012, KBB2012, khan2013}.

In this work we have studied the details of the interactions between single stars and
the MBHB with an unprecedented detail and statistical quality due to the large particle
number in our simulations (obtained with the $\varphi$GPU code on large GPU accelerated
supercomputers in China and Germany). The detailed evolution of energy and angular
momentum of the MBHB during a large number of {\sl slingshot} interactions with stars
is analyzed. Also the effect of a large scale rotation of the stellar cluster surrounding
the MBHB binary is taken into account.

The nuclear stellar cluster surrounding the MBHB is simulated with 
direct high-accuracy $N$-body simulation (Hermite scheme, \PGPU ~\citep{BNZ2011} code) with up to
$10^6$ equal mass stars. A parameter study is presented with different mass ratios 
of the black holes to each other and to the single stars (see Table~\ref{tab:mod}).
We build an efficient method to select the ejected stars from the simulations in
order to understand the detailed properties of ejected stars and how they change
the eccentricity of the MBHBs when \textit{slingshot} effects dominate the
hardening of the MBHBs. About $0.08\%$ to $8\%$ of stars are ejected by MBHBs in
our $150$ $N$-body time unit simulations (see
Table~\ref{tab:table-eject}, if we use the scale factor as discussed in the last
part of Section~\ref{sec:ui}, $150$ is about $2.25Gyr$).

Our results (see Figure~\ref{fig:bh_xyz}) exhibit two different classes of systems
based on the MBHB's rotational axis direction at the time it becomes gravitationally
bound, which we denote as $I$-model (the inclination angle
between MBHBs' orbits and stellar system rotational symmetry plane is large) and $P$-model
(MBHBs orbital plane is nearly parallel to the stellar system's rotational symmetry plane).
$I$-model and $P$-model lead to
different characteristics of the angular momentum distribution of ejected stars
(Figure~\ref{fig:ad},\ref{fig:ad_io},\ref{fig:theta_io}). The histogram reflects
both the rotation of the surrounding star cluster as well as the one of the MBHB
- there is a maximum both at angular momenta perpendicular to the orbit of MBHB,
and another one aligned with the stellar system (see I-models in
Figure~\ref{fig:theta_io}). If the stellar system is spherically symmetric we
only see the maximum at ejected stellar orbits co-rotating with the MHBH (see
last two histograms in Figure~\ref{fig:theta_io}).

Besides, the larger mass MBHBs have a stronger rotational correlation with
ejected stars. If the black hole and the stellar system's rotational symmetry plane are
similar, the effect is even stronger. For the $P$-model, the stellar system's rotational symmetry
plane dominates the concentration features of ejected stars (see
Figure~\ref{fig:ad},\ref{fig:theta_io}).

Finally, we find that massive MBHB (models $4040$ and $4020$ in
Figure~\ref{fig:theta_io}) in both rotating and non-rotating galactic nuclei
deplete co-rotating stars, because relatively more ejected stars are co-rotating
with the MBHB. This agrees with
models of
\cite{Zier2001,Zier2002,Iwasawa2011,Meiron2013}. 
\cite{Iwasawa2011} carried out simulations with small mass ratios 
(1/100) and they only considered bound stars and a non-rotating stellar 
system. They argue that non-axisymmetric perturbations by the secondary
 black hole create an effect, which is also seen in our simulations with 
rotating models containing a massive MBHB and non-rotating models 
(see Figure~\ref{fig:theta_io}): 
initially counter-rotating stars become co-rotating after being 
scattered. Our results show this effect for much more general conditions 
(up to equal mass ratio of the MBHB and including mostly unbound scattered
 stars).
The average eccentricity changes of our MBHBs agree fairly well with early Monte
Carlo prediction \citep{Q1996} (see Figure~\ref{fig:ke}), but the scatter for
individual events is quite large. This means that our interactions, detected in
the numerical simulations, cover a different, and we think more realistic range
of encounters than used in the early investigation of \cite{Q1996}.  This could
be due to different distributions of relative velocities, impact parameters or the
movement of the MBHB. This effect is more pronounced for the few cases where we
have low eccentricity, while for the large $e$ our results follow the same trend
as \cite{Q1996}.  Our data show that the eccentricity grows in a stochastic
way, where positive and negative $K$ occur all the time, but there is an average
trend towards higher eccentricity. The relativistic Post-Newtonian evolution
of the MBHB and its gravitational wave emission in the final phase before
coalescence depends on such detailed orbital evolution, which is a reason why
we need simulations like ours and others for a correct assessment of
gravitational radiation from MBHB in the universe (cf. e.g. \cite{Preto2011,khan2011,khan2013}). 
We obtain an average higher eccentricities as comparied to
\cite{khan2013}. The reason for this is probably that we have 
rotating models while their galaxy models are non-rotating.


\section*{ACKNOWLEDGMENTS}

We thank the anonymous referee for constructive comments that helped to improve the paper.

We acknowledge support by Chinese Academy of Sciences through the Silk 
Road Project at NAOC, through the Chinese Academy of Sciences Visiting 
Professorship for Senior International Scientists, Grant Number $2009S1-5$ 
(RS), and through the "Qianren" special foreign experts program of China. 

LW acknowledges support and hospitality through research visits at the 
Max-Planck Institute for Astronomy (MPA) in Garching, at the University 
of Heidelberg (SFB881), and through the European Gravitational Wave 
Observatory (EGO), VESF grant EGO-DIR-50-2010 to attend a school in 
Rome, Italy. 

The special GPU accelerated supercomputer {\tt laohu} at the Center of Information
and Computing at National Astronomical Observatories, Chinese Academy of
Sciences, funded by Ministry of Finance of People's Republic of China under the
grant $ZDYZ2008-2$, has been used for some of the largest simulations.
We also used smaller GPU clusters  {\tt titan}, {\tt hydra} and {\tt kepler}, 
funded under the grants I/80041-043 and I/84678/84680 of the Volkswagen 
Foundation and grants 823.219-439/30 and /36 of the Ministry of 
Science, Research and the Arts of Baden-W\"urttemberg, Germany. 

Some code development was also done on the Milky Way supercomputer, funded by 
the Deutsche Forschungsgemeinschaft (DFG) through Collaborative Research Center 
(SFB 881) "The Milky Way System" (subproject Z2), hosted and co-funded by the 
J\"ulich Supercomputing Center (JSC). 

PB acknowledges the special support by the NASU under the Main Astronomical
Observatory GRID/GPU computing cluster project. 

MBNK was supported by the Peter and Patricia Gruber Foundation through the PPGF fellowship, by the Peking University One Hundred Talent Fund (985), and by the National Natural Science Foundation of China (grants 11010237, 11050110414, 11173004). This publication was made possible through the support of a grant from the John Templeton Foundation and National Astronomical Observatories of Chinese Academy of Sciences. The opinions expressed in this publication are those of the author(s) do not necessarily reflect the views of the John Templeton Foundation or National Astronomical Observatories of Chinese Academy of Sciences. The funds from John Templeton Foundation were awarded in a grant to The University of Chicago which also managed the program in conjunction with National Astronomical Observatories, Chinese Academy of Sciences.


\bibliographystyle{apj}

\bibliography{apj-jour,paper}

\begin{thebibliography}{46}
\expandafter\ifx\csname natexlab\endcsname\relax\def\natexlab#1{#1}\fi

\bibitem[{{Aarseth}(2003)}]{Aarseth2003}
{Aarseth}, S.~J. 2003, \apss, 285, 367

\bibitem[{{Aarseth} {et~al.}(1974){Aarseth}, {Henon}, \& {Wielen}}]{AHW1974}
{Aarseth}, S.~J., {Henon}, M., \& {Wielen}, R. 1974, \aap, 37, 183

\bibitem[{{Amaro-Seoane} {et~al.}(2010){Amaro-Seoane}, {Sesana}, {Hoffman},
  {Benacquista}, {Eichhorn}, {Makino}, \& {Spurzem}}]{Amaro2010}
{Amaro-Seoane}, P., {Sesana}, A., {Hoffman}, L., {et~al.} 2010, \mnras, 402,
  2308

\bibitem[{{Begelman} {et~al.}(1980){Begelman}, {Blandford}, \&
  {Rees}}]{Begelman1980}
{Begelman}, M.~C., {Blandford}, R.~D., \& {Rees}, M.~J. 1980, \nat, 287, 307

\bibitem[{{Berczik} {et~al.}(2005){Berczik}, {Merritt}, \&
  {Spurzem}}]{Berczik2005}
{Berczik}, P., {Merritt}, D., \& {Spurzem}, R. 2005, \apj, 633, 680

\bibitem[{{Berczik} {et~al.}(2006){Berczik}, {Merritt}, {Spurzem}, \&
  {Bischof}}]{Berczik2006}
{Berczik}, P., {Merritt}, D., {Spurzem}, R., \& {Bischof}, H.-P. 2006, \apjl,
  642, L21

\bibitem[{{Berczik} {et~al.}(2011){Berczik}, {Nitadori}, {Zhong}, {Spurzem},
  {Hamada}, {Berentzen}, \& {Veles}}]{BNZ2011}
{Berczik}, P., {Nitadori}, K., {Zhong}, S., {et~al.} 2011, in High Performence
  Computing HPC-UA, Vol.~4, 8

\bibitem[{{Berczik} {et~al.}(2013{\natexlab{a}}){Berczik}, {Spurzem},
  {Berentzen}, \& {Nitadori}}]{BSBN2013}
{Berczik}, P., {Spurzem}, R., {Berentzen}, I., \& {Nitadori}, K.
  2013{\natexlab{a}}, \apj, will be submitted

\bibitem[{{Berczik} {et~al.}(2013{\natexlab{b}}){Berczik}, {Spurzem}, {Zhong},
  {Wang}, {Nitadori}, {Hamada}, \& {Veles}}]{Berczik2013}
{Berczik}, P., {Spurzem}, R., {Zhong}, S., {et~al.} 2013{\natexlab{b}}, in
  Lecture Notes in Computer Science, Vol. 7905, Procs. of 28th Intl.
  Supercomputing Conf. ISC 2013, Leipzig, Germany, June 16-20, 2013., ed. J.~M.
  {Kunkel}, T.~{Ludwig}, \& H.~E. {Meuer} (Springer Vlg.), 13--25

\bibitem[{{Berentzen} {et~al.}(2008){Berentzen}, {Preto}, {Berczik}, {Merritt},
  \& {Spurzem}}]{BPB2008}
{Berentzen}, I., {Preto}, M., {Berczik}, P., {Merritt}, D., \& {Spurzem}, R.
  2008, Astronomische Nachrichten, 329, 904

\bibitem[{{Berentzen} {et~al.}(2009{\natexlab{a}}){Berentzen}, {Preto},
  {Berczik}, {Merritt}, \& {Spurzem}}]{Berentzen2009}
---. 2009{\natexlab{a}}, \apj, 695, 455

\bibitem[{{Berentzen} {et~al.}(2009{\natexlab{b}}){Berentzen}, {Preto},
  {Berczik}, {Merritt}, \& {Spurzem}}]{BPBMS2009}
---. 2009{\natexlab{b}}, \apj, 695, 455

\bibitem[{{Einsel} \& {Spurzem}(1999)}]{ES1999}
{Einsel}, C., \& {Spurzem}, R. 1999, \mnras, 302, 81

\bibitem[{{Fiestas} {et~al.}(2012){Fiestas}, {Porth}, {Berczik}, \&
  {Spurzem}}]{FPB2012}
{Fiestas}, J., {Porth}, O., {Berczik}, P., \& {Spurzem}, R. 2012, \mnras, 419,
  57

\bibitem[{{Fukushige} {et~al.}(2005){Fukushige}, {Makino}, \&
  {Kawai}}]{FMK2005}
{Fukushige}, T., {Makino}, J., \& {Kawai}, A. 2005, \pasj, 57, 1009

\bibitem[{{Gong} {et~al.}(2011){Gong}, {Xu}, {Bai}, {Cao}, {Chen}, {Chen},
  {He}, {Heinzel}, {Lau}, {Liu}, {Luo}, {Luo}, {Pulido Pat{\'o}n},
  {R{\"u}diger}, {Shao}, {Spurzem}, {Wang}, {Xu}, {Yeh}, {Yuan}, \&
  {Zhou}}]{Gong2011}
{Gong}, X., {Xu}, S., {Bai}, S., {et~al.} 2011, Classical and Quantum Gravity,
  28, 094012

\bibitem[{{Greene} \& {Ho}(2009)}]{Greene2009}
{Greene}, J.~E., \& {Ho}, L.~C. 2009, \pasp, 121, 1167

\bibitem[{{Gualandris} {et~al.}(2012){Gualandris}, {Dotti}, \&
  {Sesana}}]{GDS2012}
{Gualandris}, A., {Dotti}, M., \& {Sesana}, A. 2012, \mnras, 420, L38

\bibitem[{{Gualandris} \& {Merritt}(2012)}]{Gualandris2012}
{Gualandris}, A., \& {Merritt}, D. 2012, \apj, 744, 74

\bibitem[{{Harfst} {et~al.}(2007){Harfst}, {Gualandris}, {Merritt}, {Spurzem},
  S., \& {Berczik}}]{HGM2007}
{Harfst}, S., {Gualandris}, A., {Merritt}, D., {et~al.} 2007, \na, 12, 357

\bibitem[{{Iwasawa} {et~al.}(2011){Iwasawa}, {An}, {Matsubayashi}, {Funato}, \&
  {Makino}}]{Iwasawa2011}
{Iwasawa}, M., {An}, S., {Matsubayashi}, T., {Funato}, Y., \& {Makino}, J.
  2011, \apjl, 731, L9

\bibitem[{{Khan} {et~al.}(2013){Khan}, {Holley-Bockelmann}, {Berczik}, \&
  {Just}}]{khan2013}
{Khan}, F., {Holley-Bockelmann}, K., {Berczik}, P., \& {Just}, A. 2013, \apj,
  773, 100

\bibitem[{{Khan} {et~al.}(2012{\natexlab{a}}){Khan}, {Berentzen}, {Berczik},
  {Just}, {Mayer}, {Nitadori}, \& {Callegari}}]{KBB2012}
{Khan}, F.~M., {Berentzen}, I., {Berczik}, P., {et~al.} 2012{\natexlab{a}},
  \apj, 756, 30

\bibitem[{{Khan} {et~al.}(2011){Khan}, {Just}, \& {Merritt}}]{khan2011}
{Khan}, F.~M., {Just}, A., \& {Merritt}, D. 2011, \apj, 732, 89

\bibitem[{{Khan} {et~al.}(2012{\natexlab{b}}){Khan}, {Preto}, {Berczik},
  {Berentzen}, {Just}, \& {Spurzem}}]{KPBB2012}
{Khan}, F.~M., {Preto}, M., {Berczik}, P., {et~al.} 2012{\natexlab{b}}, \apj,
  749, 147

\bibitem[{{Li} {et~al.}(2012){Li}, {Liu}, {Berczik}, {Chen}, \&
  {Spurzem}}]{LLB2012}
{Li}, S., {Liu}, F.~K., {Berczik}, P., {Chen}, X., \& {Spurzem}, R. 2012, \apj,
  748, 65

\bibitem[{{Lu} {et~al.}(2010){Lu}, {Zhang}, \& {Yu}}]{Lu2010}
{Lu}, Y., {Zhang}, F., \& {Yu}, Q. 2010, \apj, 709, 1356

\bibitem[{{Meiron} \& {Laor}(2013)}]{Meiron2013}
{Meiron}, Y., \& {Laor}, A. 2013, ArXiv e-prints

\bibitem[{{Merritt}(2001)}]{Merritt2001}
{Merritt}, D. 2001, \apj, 556, 245

\bibitem[{{Merritt} \& {Poon}(2004)}]{Merritt2004}
{Merritt}, D., \& {Poon}, M.~Y. 2004, \apj, 606, 788

\bibitem[{{Milosavljevi{\'c}} \& {Merritt}(2003)}]{Milosavlijevic2003}
{Milosavljevi{\'c}}, M., \& {Merritt}, D. 2003, \apj, 596, 860

\bibitem[{{Nitadori} \& {Makino}(2008)}]{NM2008}
{Nitadori}, K., \& {Makino}, J. 2008, \na, 13, 498

\bibitem[{{Peters}(1964)}]{Peters1964}
{Peters}, P.~C. 1964, Physical Review, 136, 1224

\bibitem[{{Preto} {et~al.}(2009){Preto}, {Berentzen}, {Berczik}, {Merritt}, \&
  {Spurzem}}]{Preto2009}
{Preto}, M., {Berentzen}, I., {Berczik}, P., {Merritt}, D., \& {Spurzem}, R.
  2009, Journal of Physics Conference Series, 154, 012049

\bibitem[{{Preto} {et~al.}(2011){Preto}, {Berentzen}, {Berczik}, \&
  {Spurzem}}]{Preto2011}
{Preto}, M., {Berentzen}, I., {Berczik}, P., \& {Spurzem}, R. 2011, \apjl, 732,
  L26

\bibitem[{{Quinlan}(1996)}]{Q1996}
{Quinlan}, G.~D. 1996, \na, 1, 35

\bibitem[{{Quinlan} \& {Hernquist}(1997)}]{Quinlan1997}
{Quinlan}, G.~D., \& {Hernquist}, L. 1997, \na, 2, 533

\bibitem[{{Sesana} {et~al.}(2011){Sesana}, {Gualandris}, \&
  {Dotti}}]{Sesana2011}
{Sesana}, A., {Gualandris}, A., \& {Dotti}, M. 2011, \mnras, 415, L35

\bibitem[{{Spurzem} {et~al.}(2011{\natexlab{a}}){Spurzem}, {Berczik},
  {Berentzen}, {Ge}, {Wang}, {Schive}, {Nitadori}, \& {Hamada}}]{spurzem2011a}
{Spurzem}, R., {Berczik}, P., {Berentzen}, I., {et~al.} 2011{\natexlab{a}}, in
  Large Scale Computing Techniques for Complex Systems and Simulations, ed.
  W.~{Dubitzky}, K.~{Kurowski}, \& B.~{Schott}, Wiley Publishers, 35--58

\bibitem[{{Spurzem} {et~al.}(2012){Spurzem}, {Berczik}, {Zhong}, {Nitadori},
  {Hamada}, {Berentzen}, \& {Veles}}]{SBZ2012}
{Spurzem}, R., {Berczik}, P., {Zhong}, S., {et~al.} 2012, in Astronomical
  Society of the Pacific Conference Series, Vol. 453, Advances in Computational
  Astrophysics: Methods, Tools, and Outcome, ed. R.~{Capuzzo-Dolcetta},
  M.~{Limongi}, \& A.~{Tornamb{\`e}}, 223

\bibitem[{{Spurzem} {et~al.}(2011{\natexlab{b}}){Spurzem}, {Berczik}, {Hamada},
  {Nitadori}, {Marcus}, {Kugel}, {M{\"a}nner}, {Berentzen}, {Fiestas},
  {Banerjee}, \& {Klessen}}]{spurzem2011b}
{Spurzem}, R., {Berczik}, P., {Hamada}, T., {et~al.} 2011{\natexlab{b}},
  Computer Science - Research and Development (CSRD), 26, 145

\bibitem[{{Volonteri} {et~al.}(2003){Volonteri}, {Haardt}, \&
  {Madau}}]{Volonteri2003}
{Volonteri}, M., {Haardt}, F., \& {Madau}, P. 2003, \apj, 582, 559

\bibitem[{{Yu}(2002)}]{Yu2002}
{Yu}, Q. 2002, \mnras, 331, 935

\bibitem[{{Yu} \& {Tremaine}(2003)}]{Yu2003}
{Yu}, Q., \& {Tremaine}, S. 2003, \apj, 599, 1129

\bibitem[{{Zier} \& {Biermann}(2001)}]{Zier2001}
{Zier}, C., \& {Biermann}, P.~L. 2001, \aap, 377, 23

\bibitem[{{Zier} \& {Biermann}(2002)}]{Zier2002}
---. 2002, \aap, 396, 91

\end{thebibliography}

\newpage



\addtocounter{figure}{-1}

\begin{figure*}[htbp]
  \begin{center}
      \begin{tabular}{c c}  
      \includegraphics[angle=0,width=1.0\columnwidth]{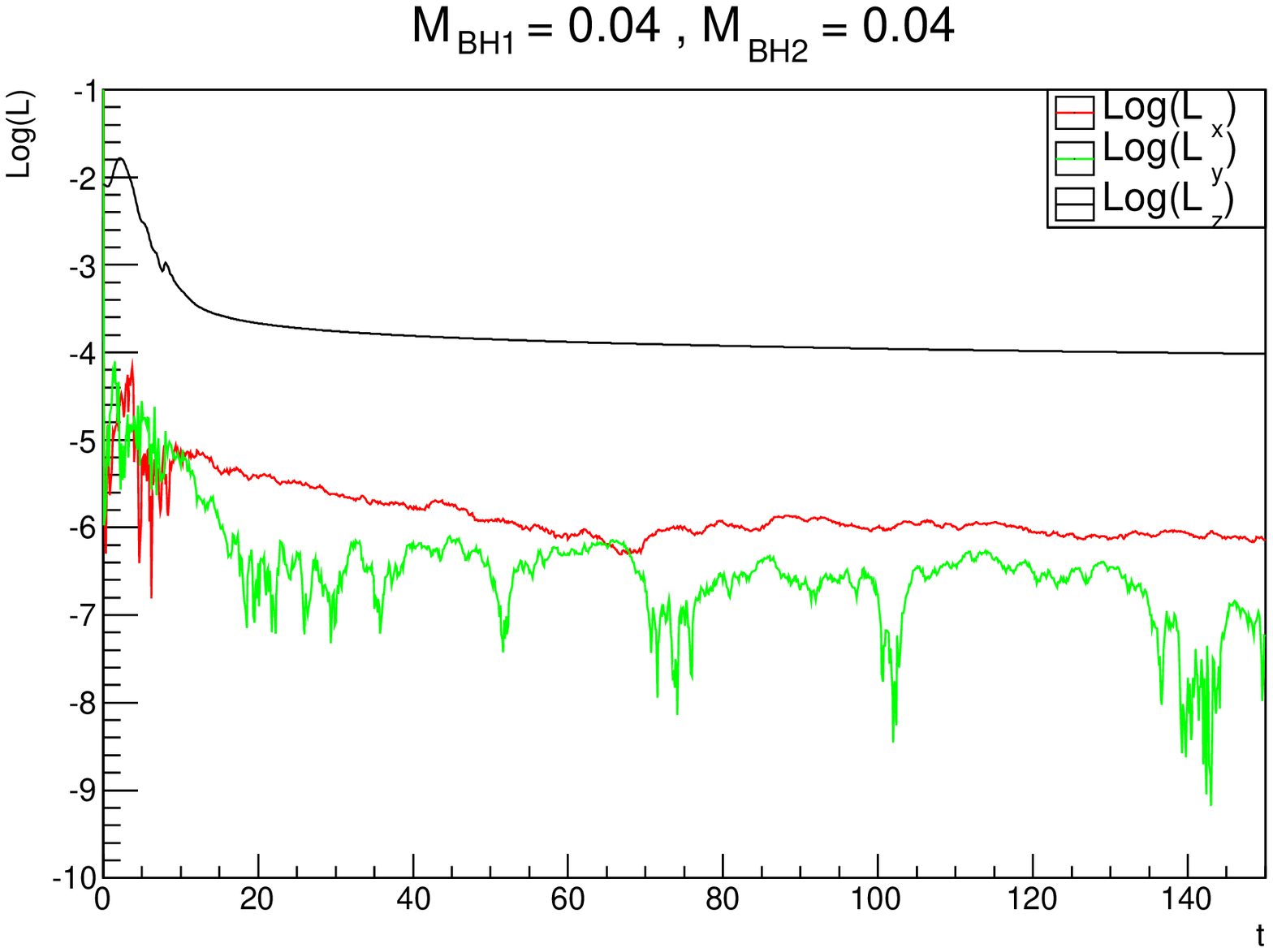}\\
      \includegraphics[angle=0,width=1.0\columnwidth]{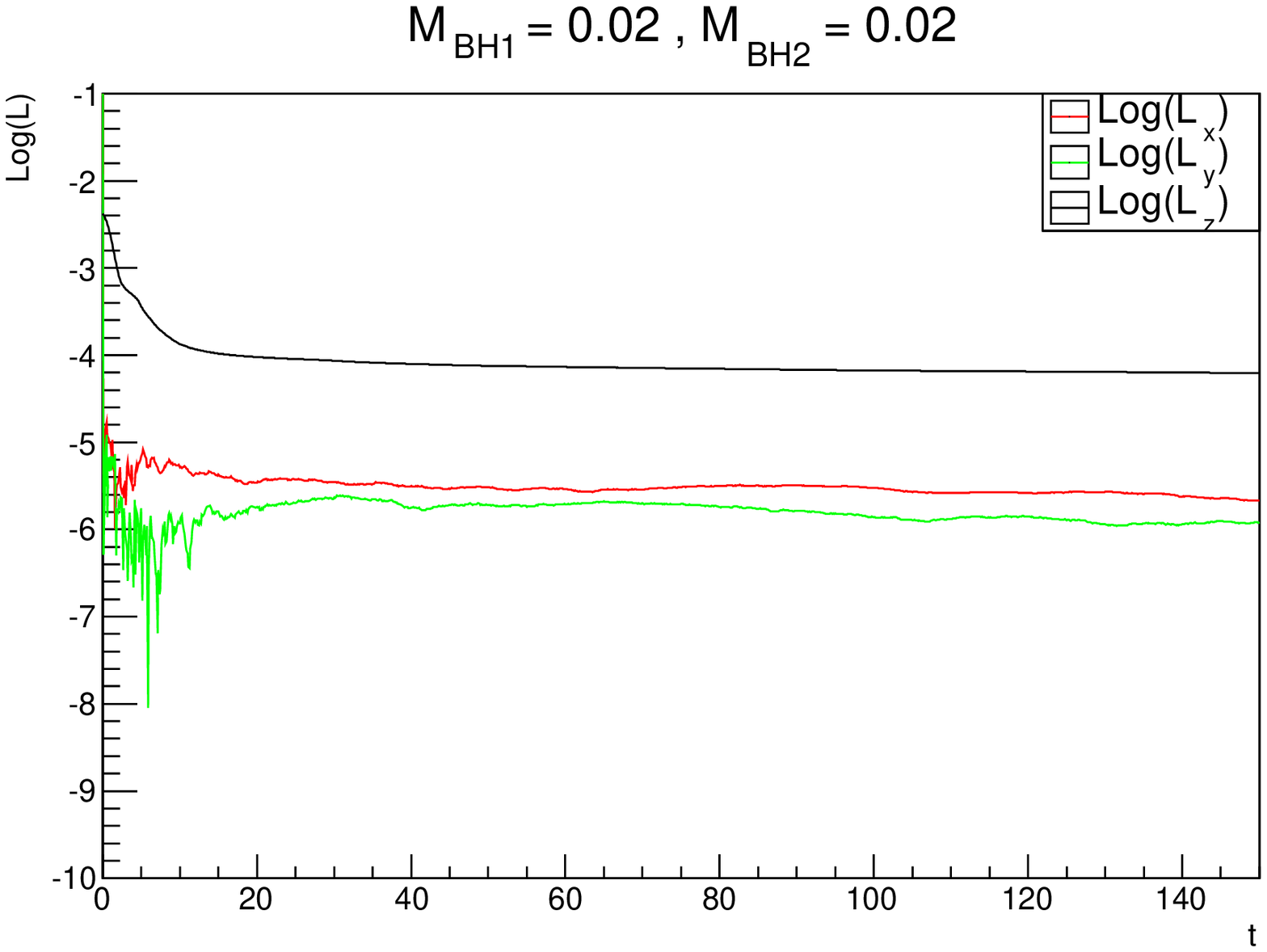}&
      \includegraphics[angle=0,width=1.0\columnwidth]{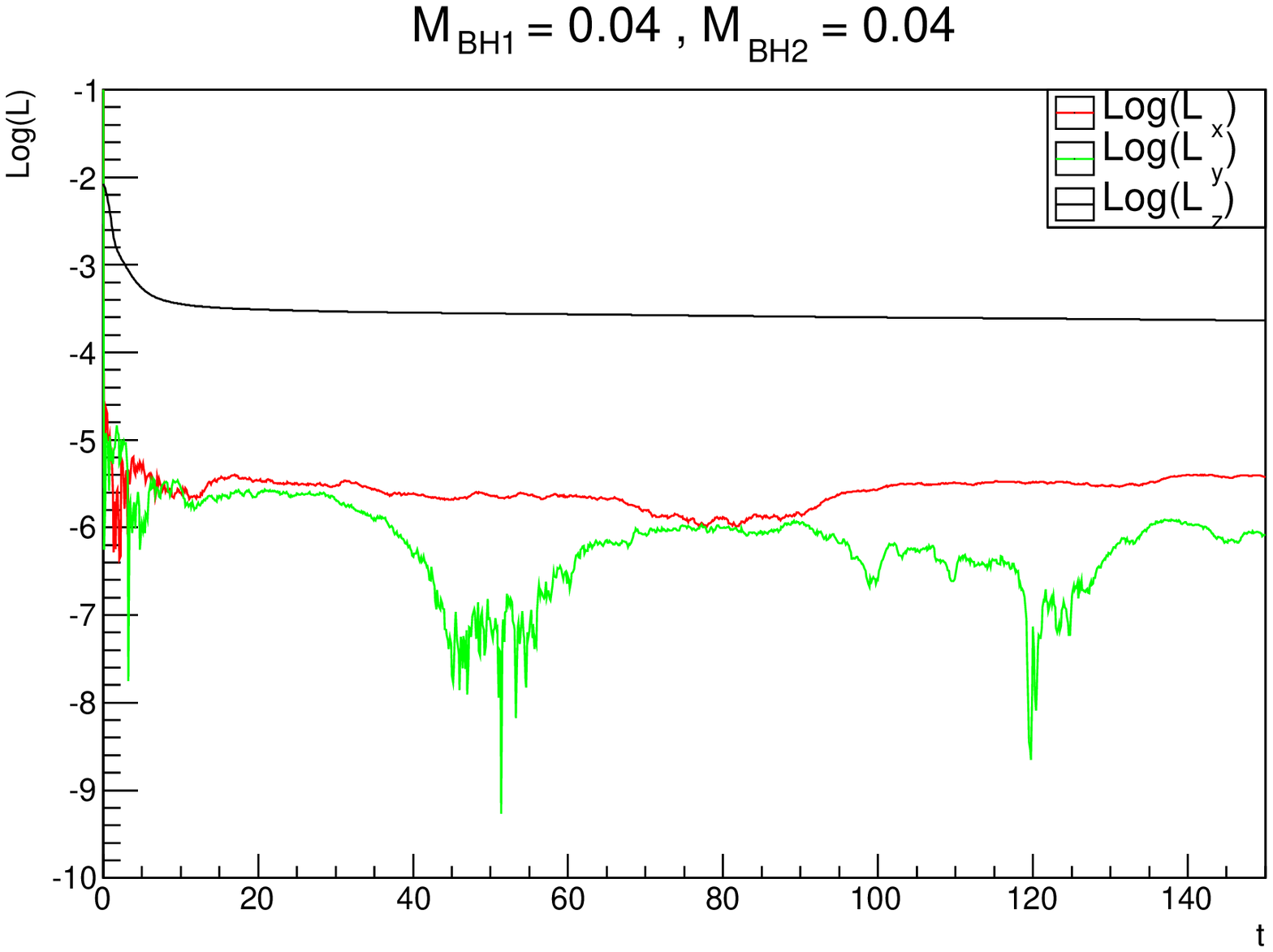}
      \end{tabular}
  \end{center}
  \caption{Continue of Figure~\ref{fig:bh_xyz}}
  \label{fig:bh_xyz:2}
\end{figure*}

 \begin{figure*}[htbp]
   \begin{center}
     \begin{tabular}{c c}
       \includegraphics[angle=0,width=1.0\columnwidth]{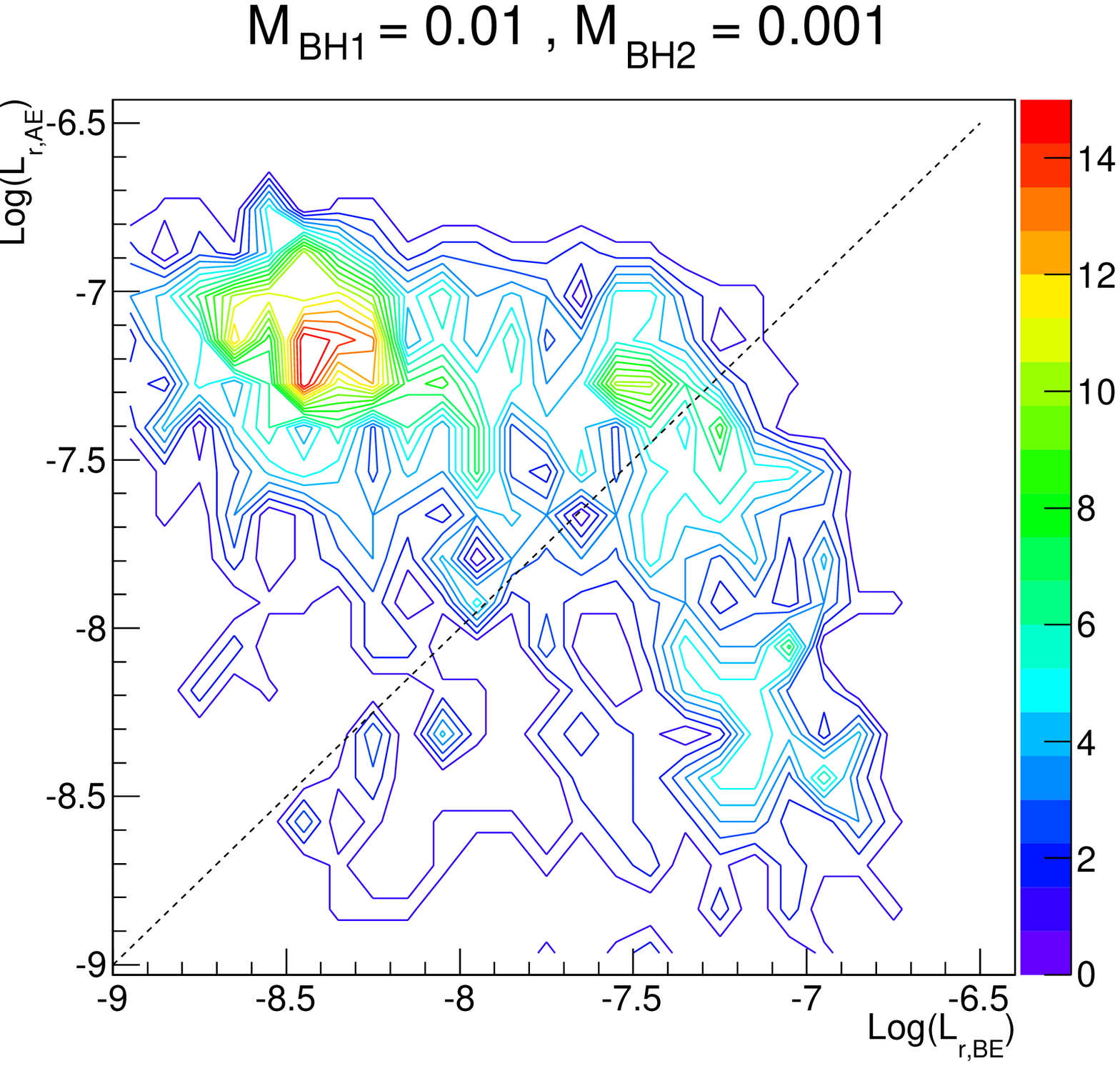}&
       \includegraphics[angle=0,width=1.0\columnwidth]{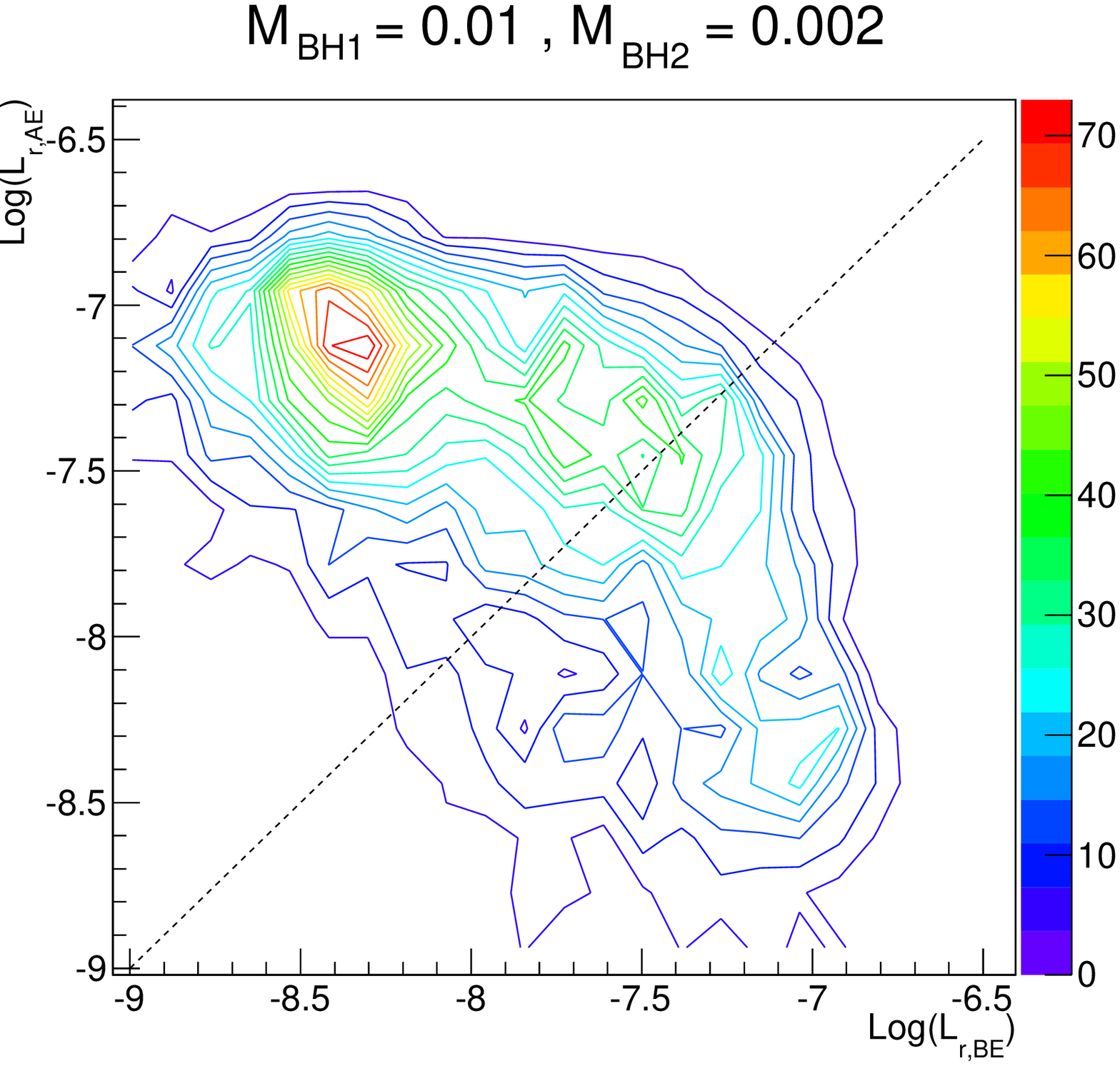}\\
       \includegraphics[angle=0,width=1.0\columnwidth]{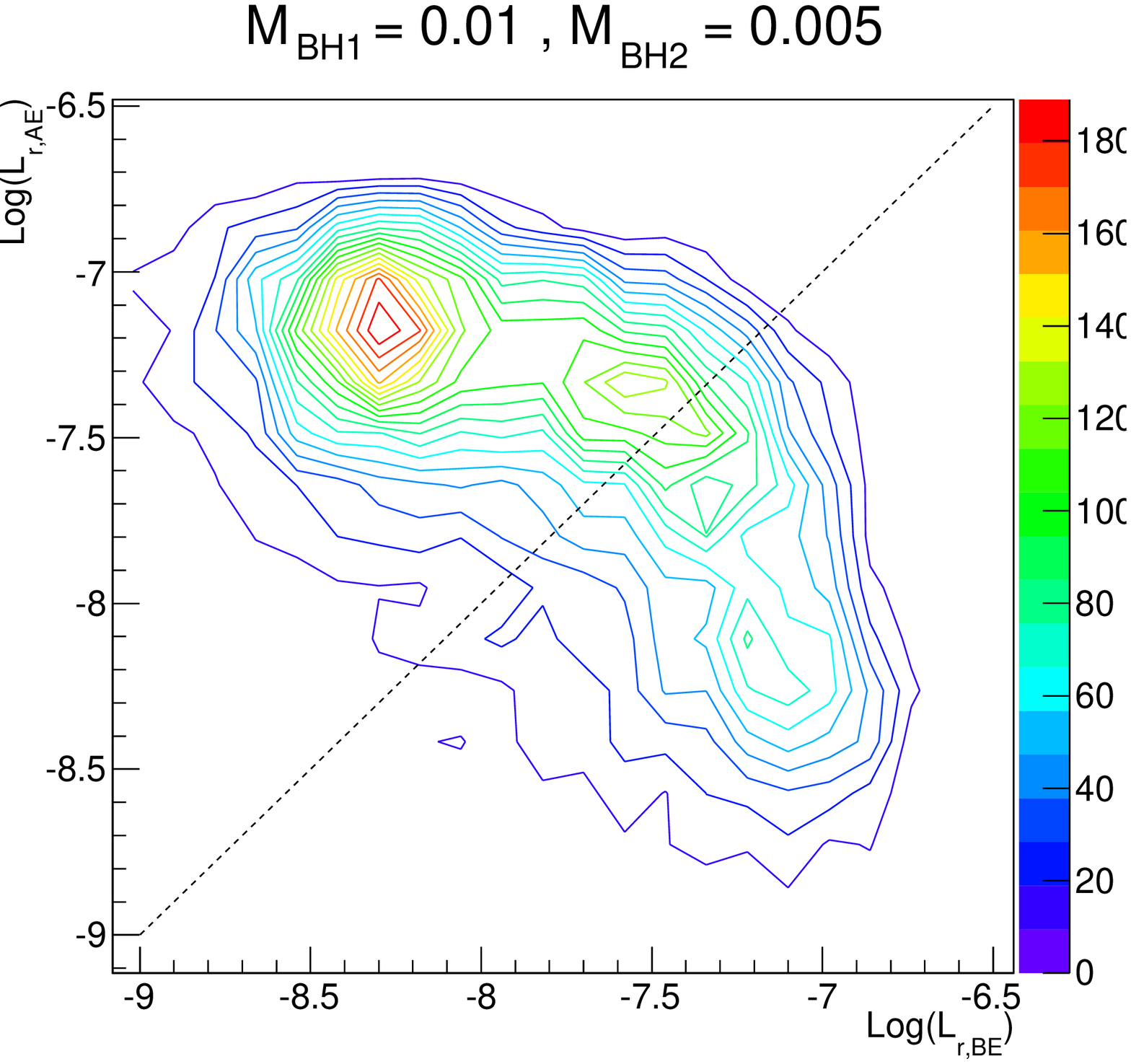}&
       \includegraphics[angle=0,width=1.0\columnwidth]{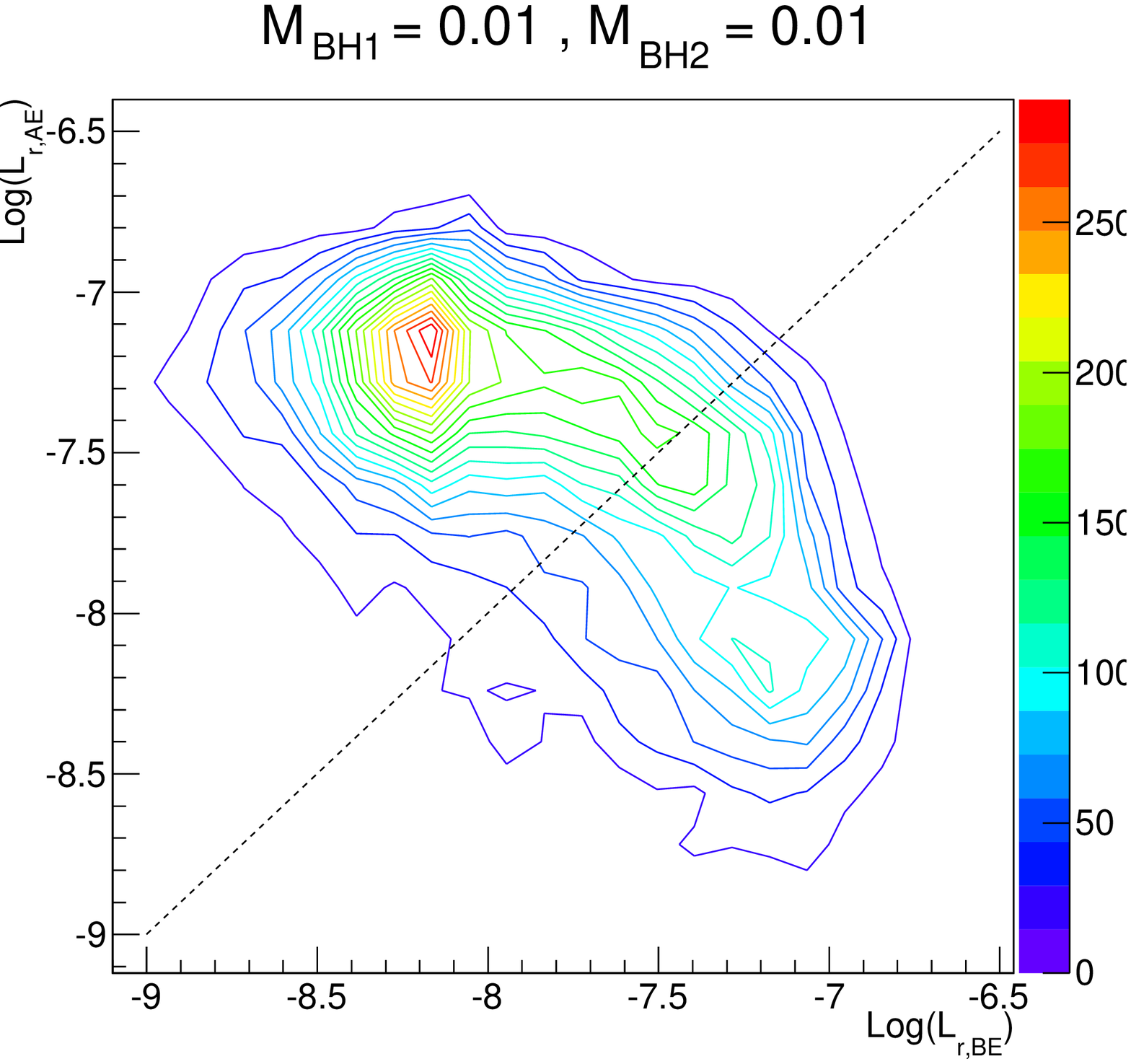}\\
     \end{tabular}
   \end{center}
   \caption{Density maps of logarithmic value of angular momentum of ejected stars
     ($\log L_r$) after vs. before ejection for all models with $t>50$ $N$-body
     time units. Along the dotted line, $L_r$ has no change before and after
     ejection. The last two panels are non-rotation models for comparison.}
   \label{fig:L_io}
 \end{figure*}
\addtocounter{figure}{-1}
 \begin{figure*}[htbp]
   \begin{center}
     \begin{tabular}{c c}
       \includegraphics[angle=0,width=1.0\columnwidth]{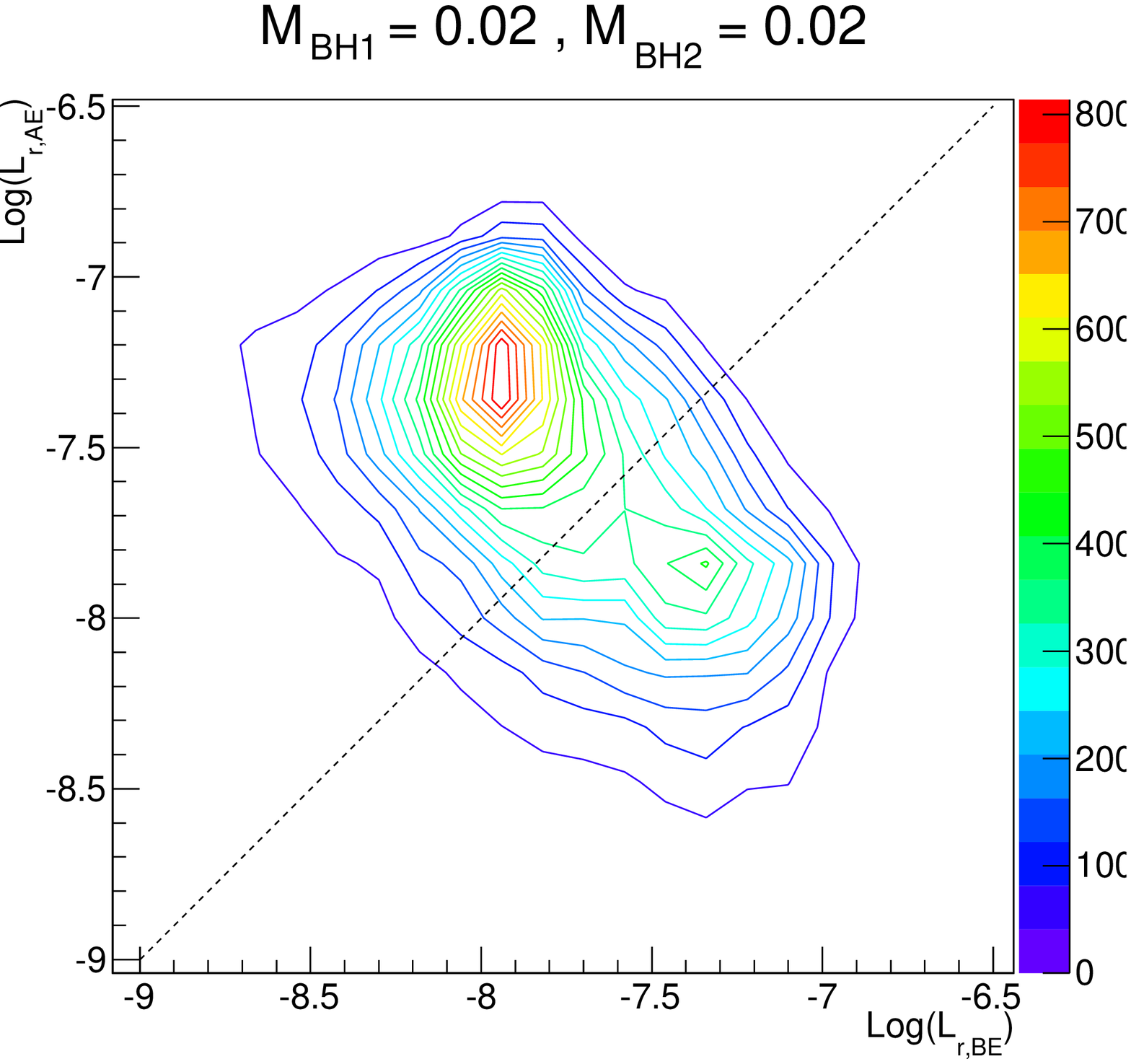}&
       \includegraphics[angle=0,width=1.0\columnwidth]{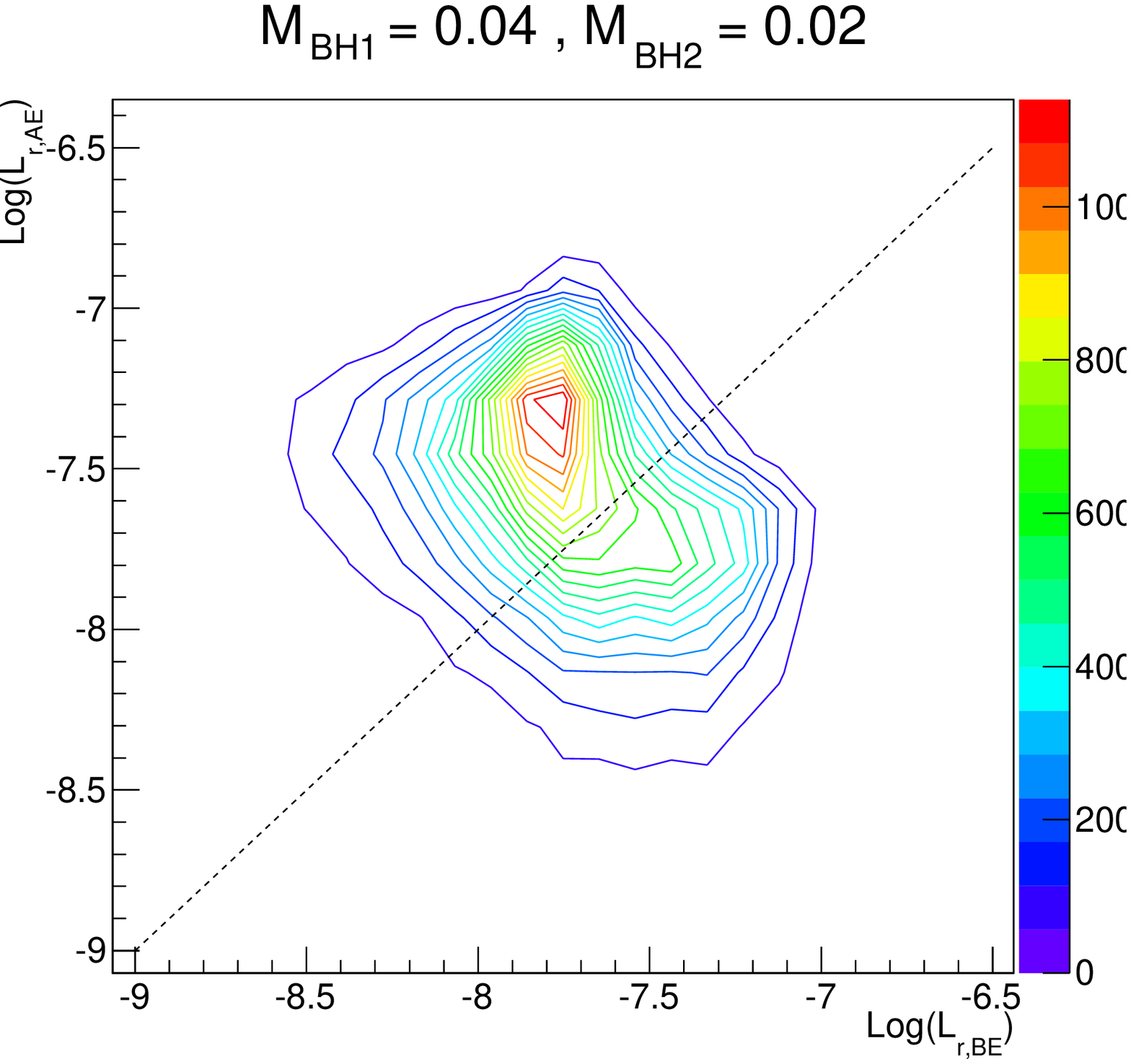}\\
       \includegraphics[angle=0,width=1.0\columnwidth]{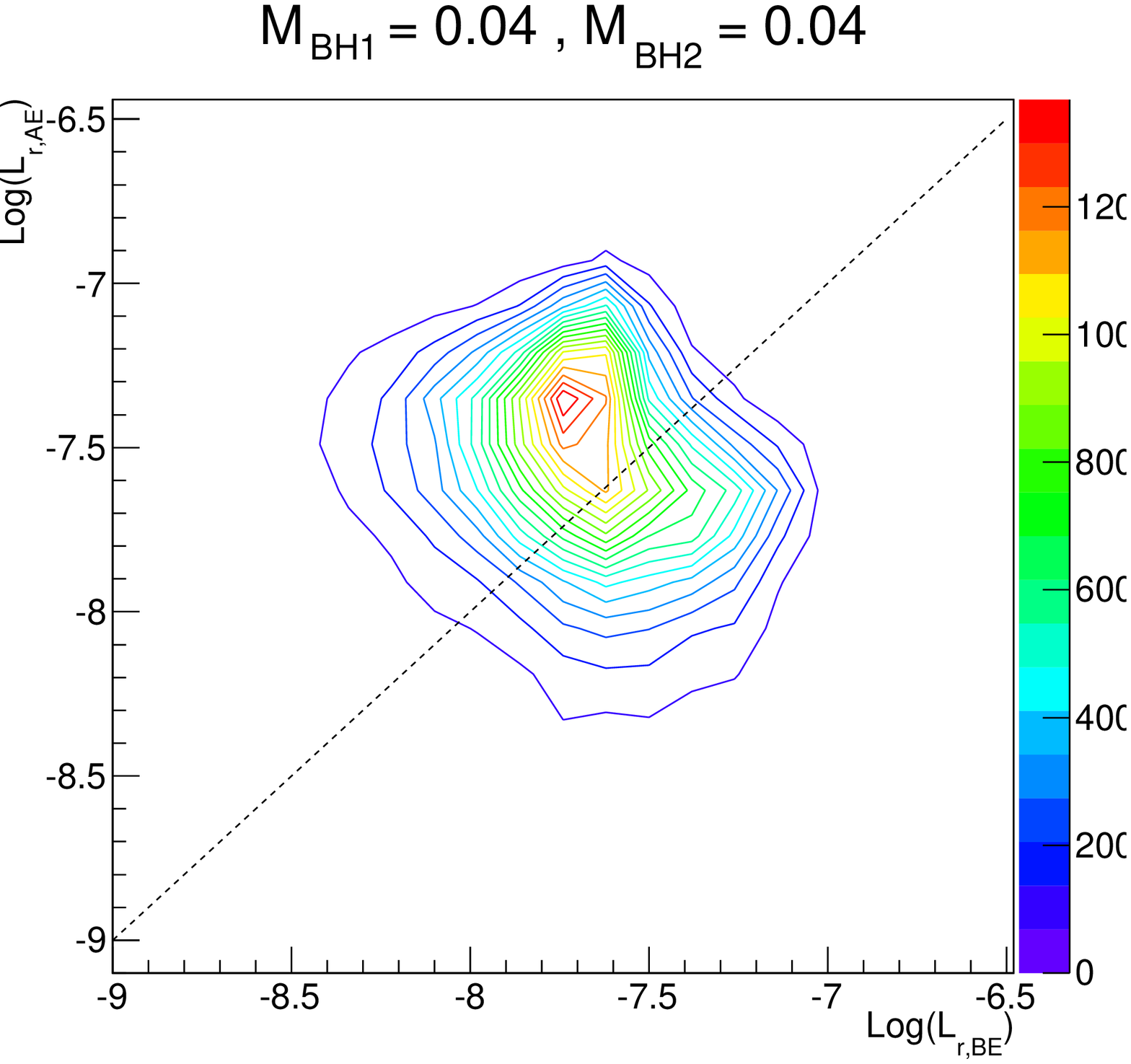}&\\
       \includegraphics[angle=0,width=1.0\columnwidth]{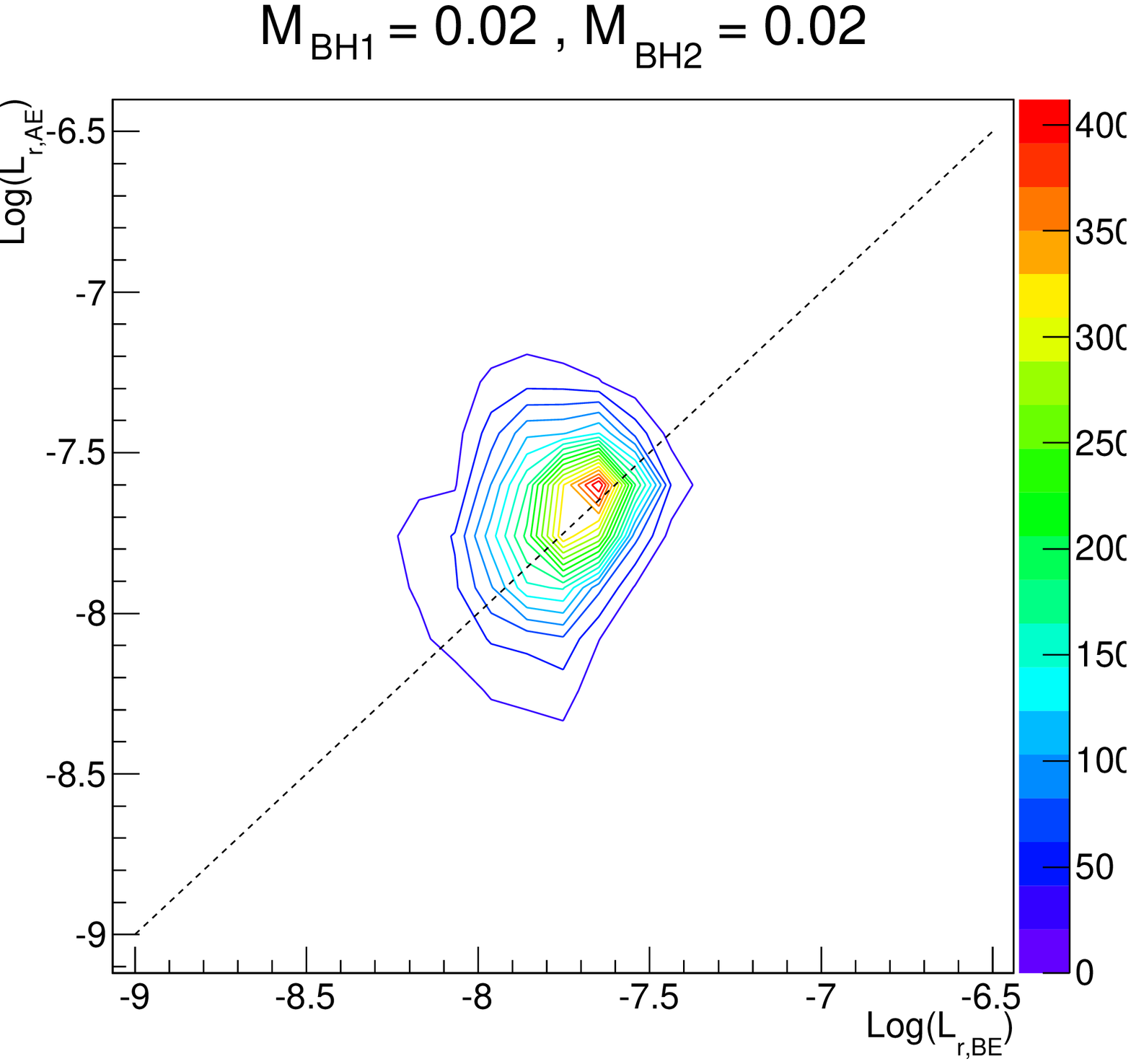}&
       \includegraphics[angle=0,width=1.0\columnwidth]{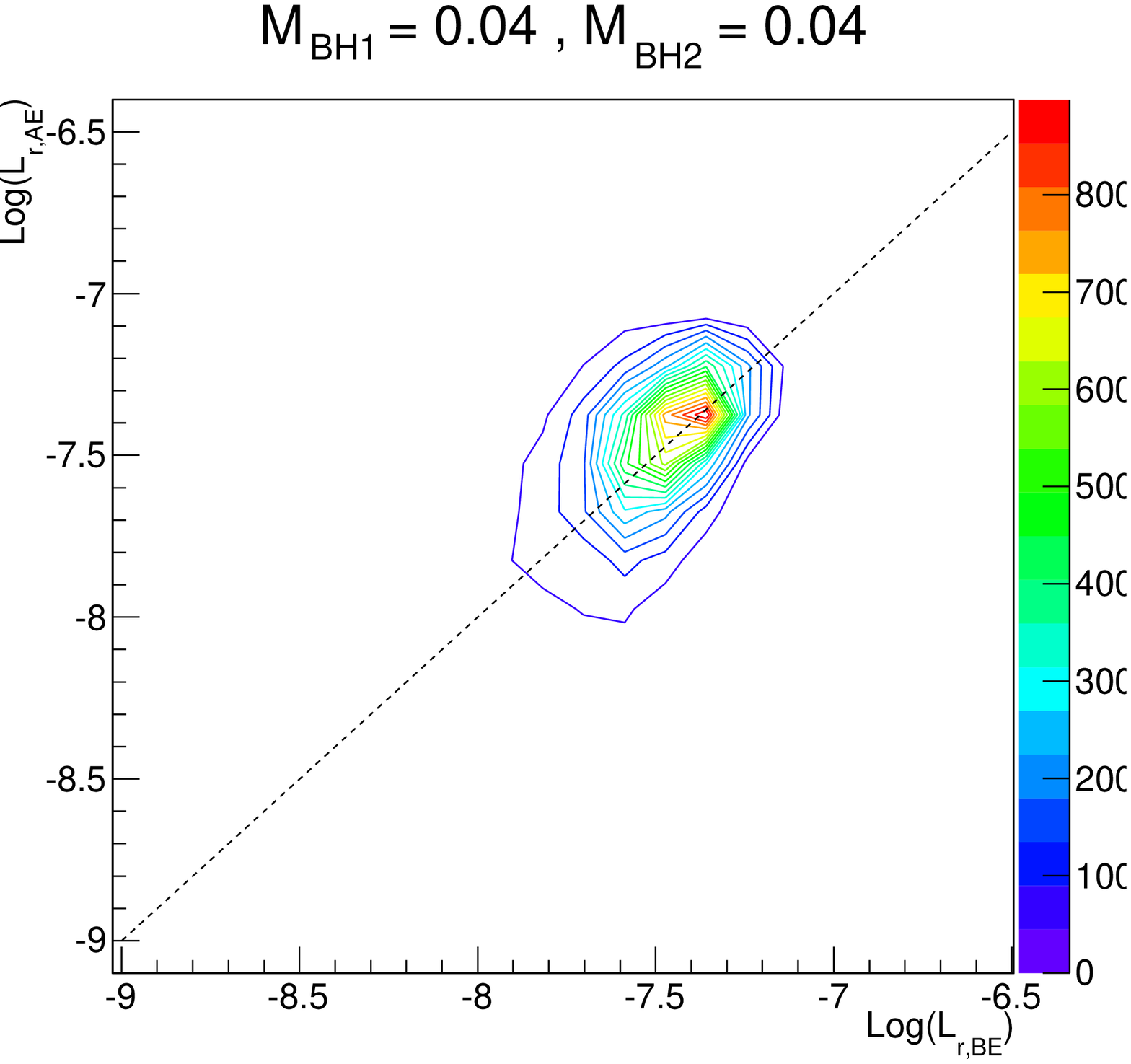}\\
     \end{tabular}
   \end{center}
   \caption{Continue of Figure~\ref{fig:L_io}}
   \label{fig:L_io:2}
 \end{figure*}

 \begin{figure*}[htbp]
   \begin{center}
     \begin{tabular}{c c}
       \includegraphics[angle=0,width=1.0\columnwidth]{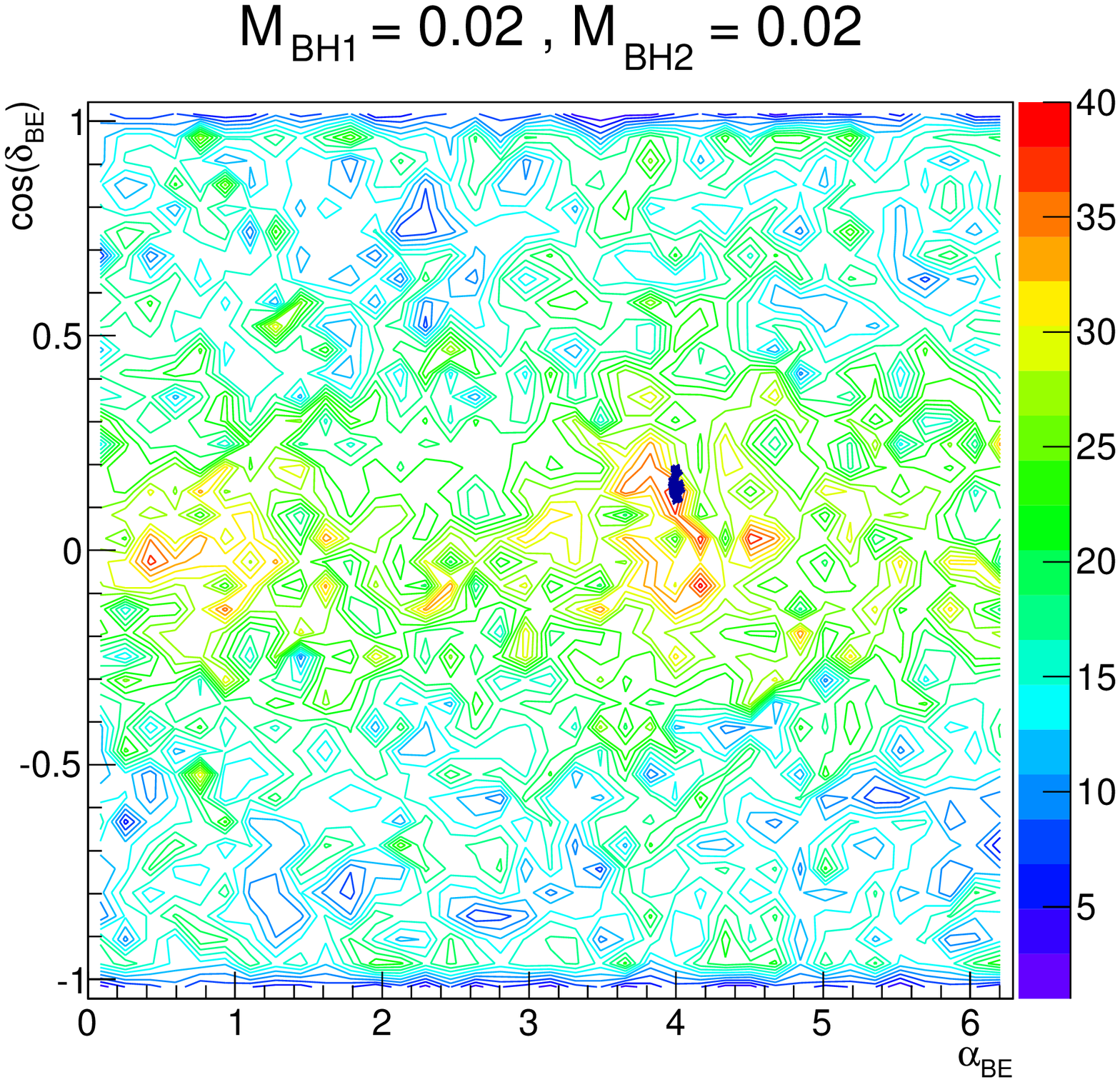}&
       \includegraphics[angle=0,width=1.0\columnwidth]{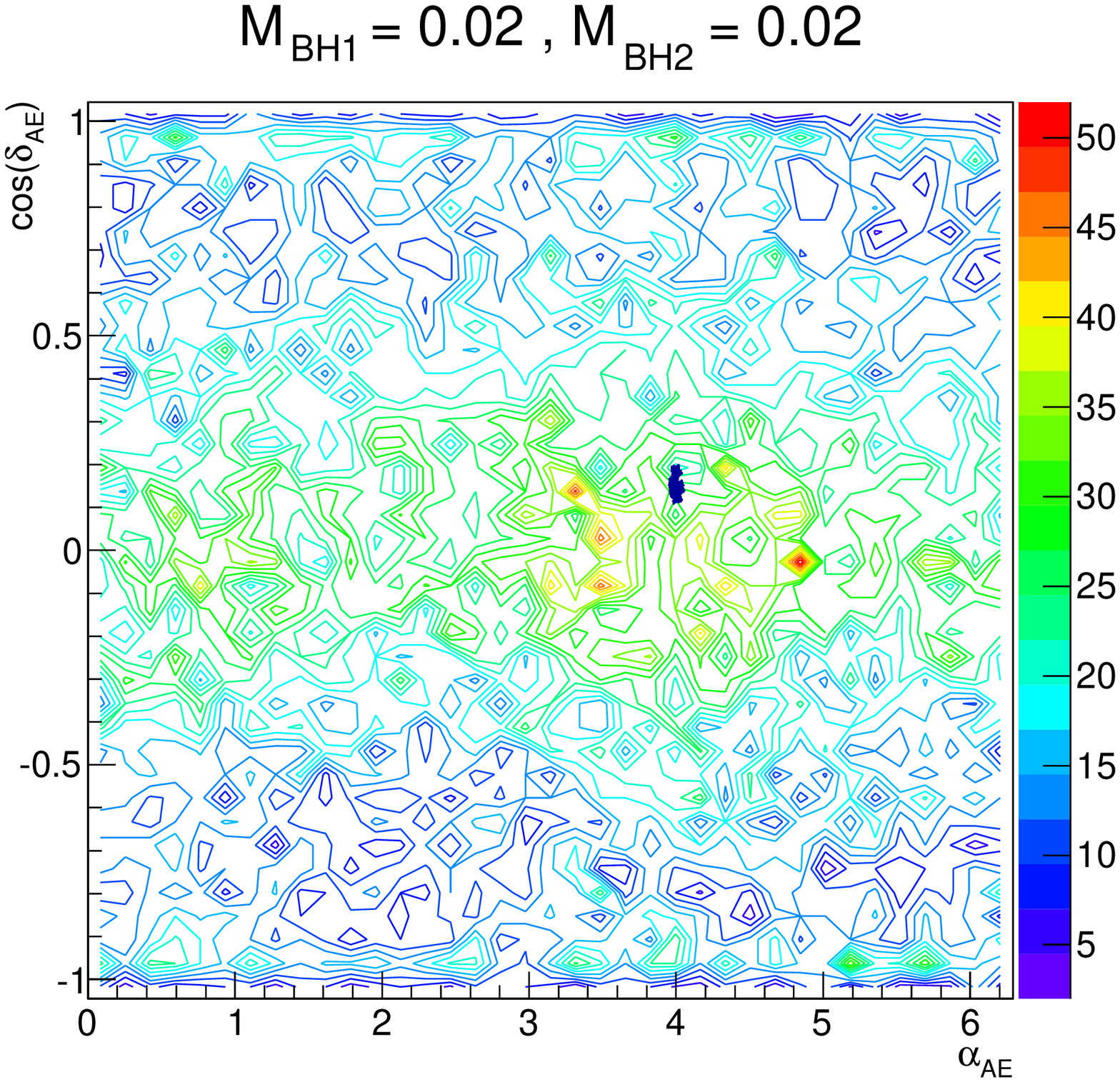}\\
       \includegraphics[angle=0,width=1.0\columnwidth]{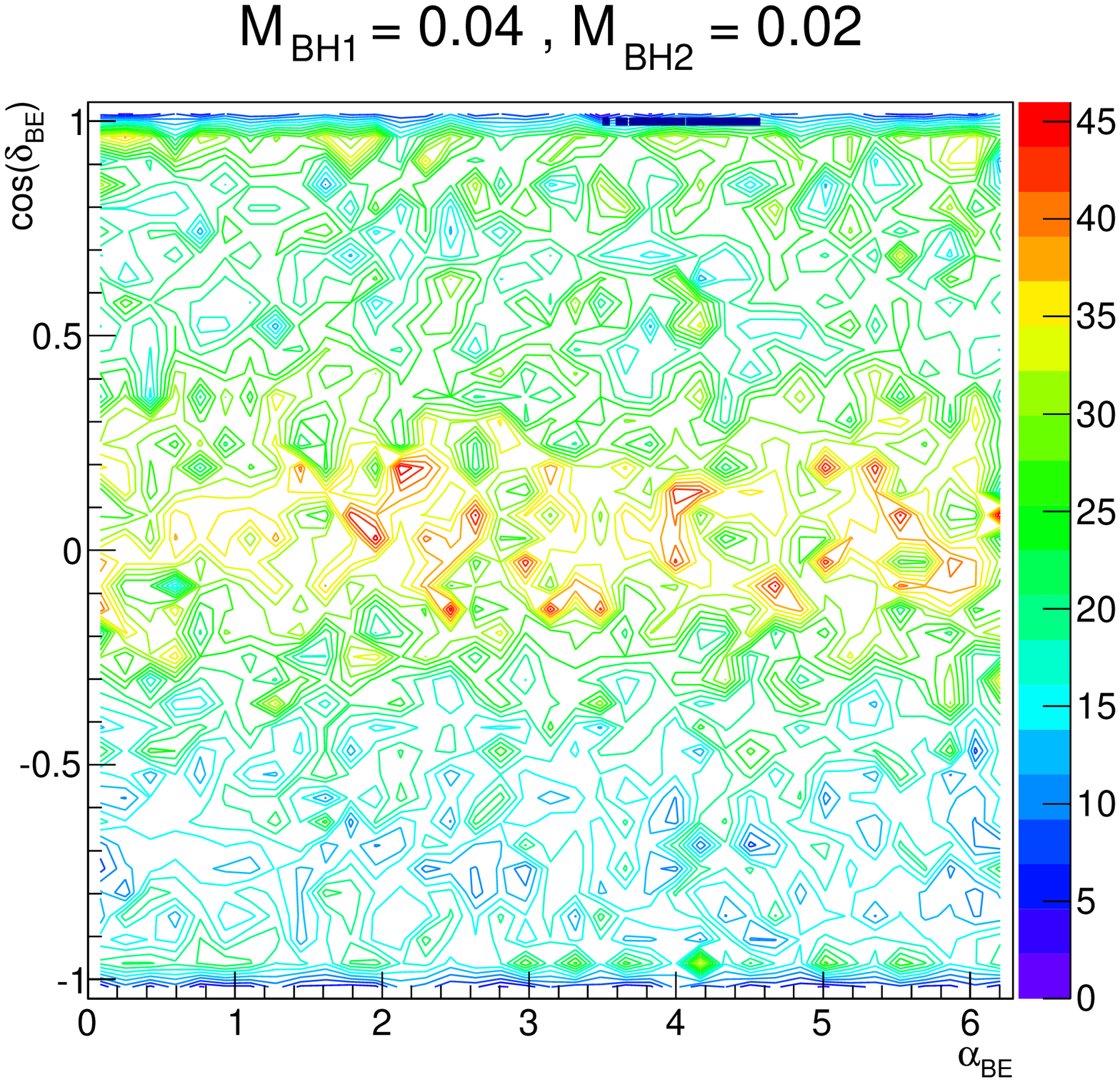}&
       \includegraphics[angle=0,width=1.0\columnwidth]{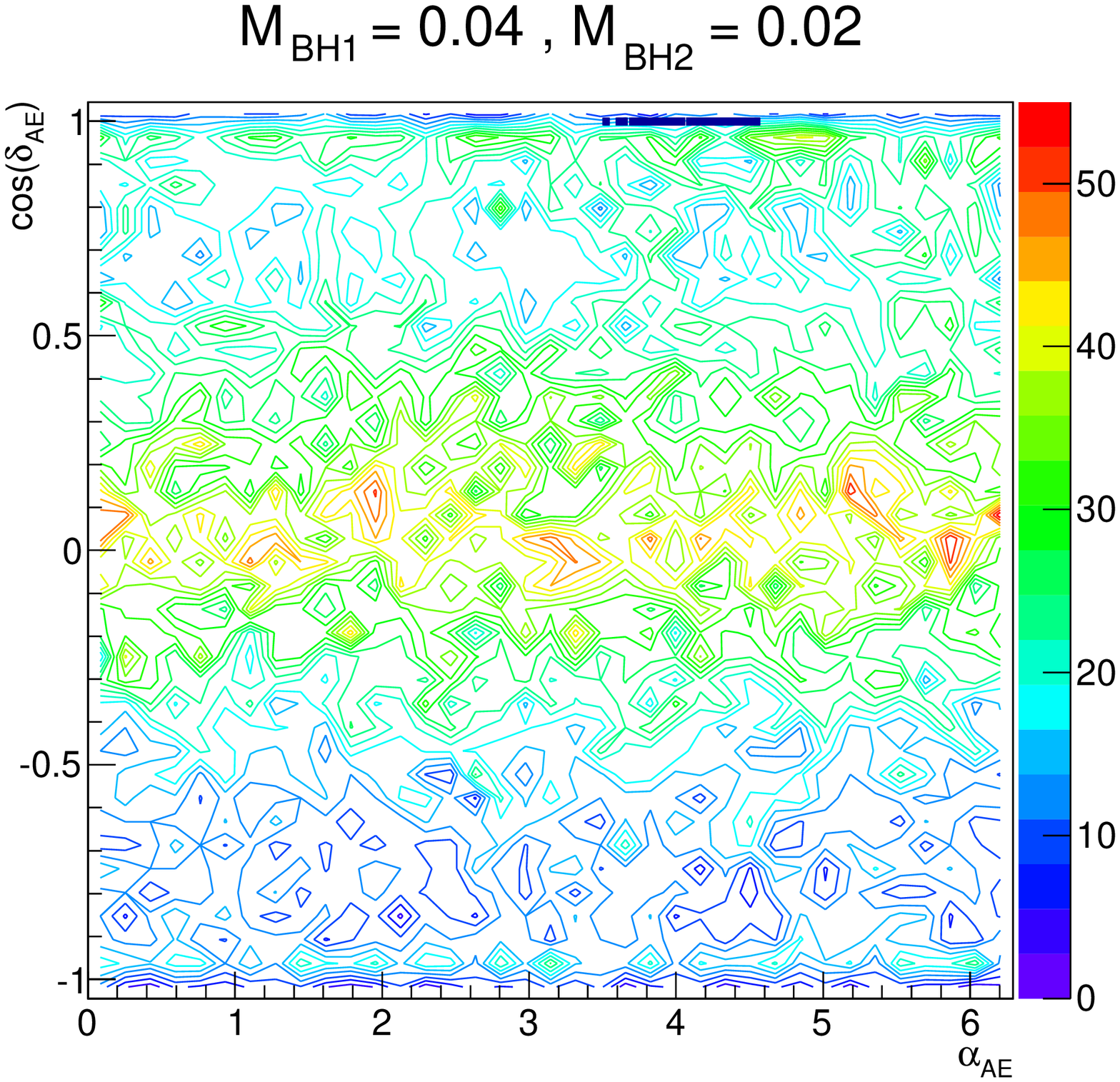}\\
      \end{tabular}
   \end{center}
   \caption{Density maps of $\delta$ vs. $\alpha$ of ejected stars and MBHBs
     with $t>50$ $N$-body time unit before (left) and after (right) ejection The
     top two panels represent model $2020$ (I-model) and the bottom two panels represent model $4020$
     (P-model). The colors of the contours represent ejected stars and black points and
     lines (near center in top panels and near top edge in the bottom panels )
     represent the MBHBs.}
   \label{fig:ad}
 \end{figure*}

 \begin{figure*}
   \begin{center}
     \begin{tabular}{c c}
       \includegraphics[angle=0,width=1.0\columnwidth]{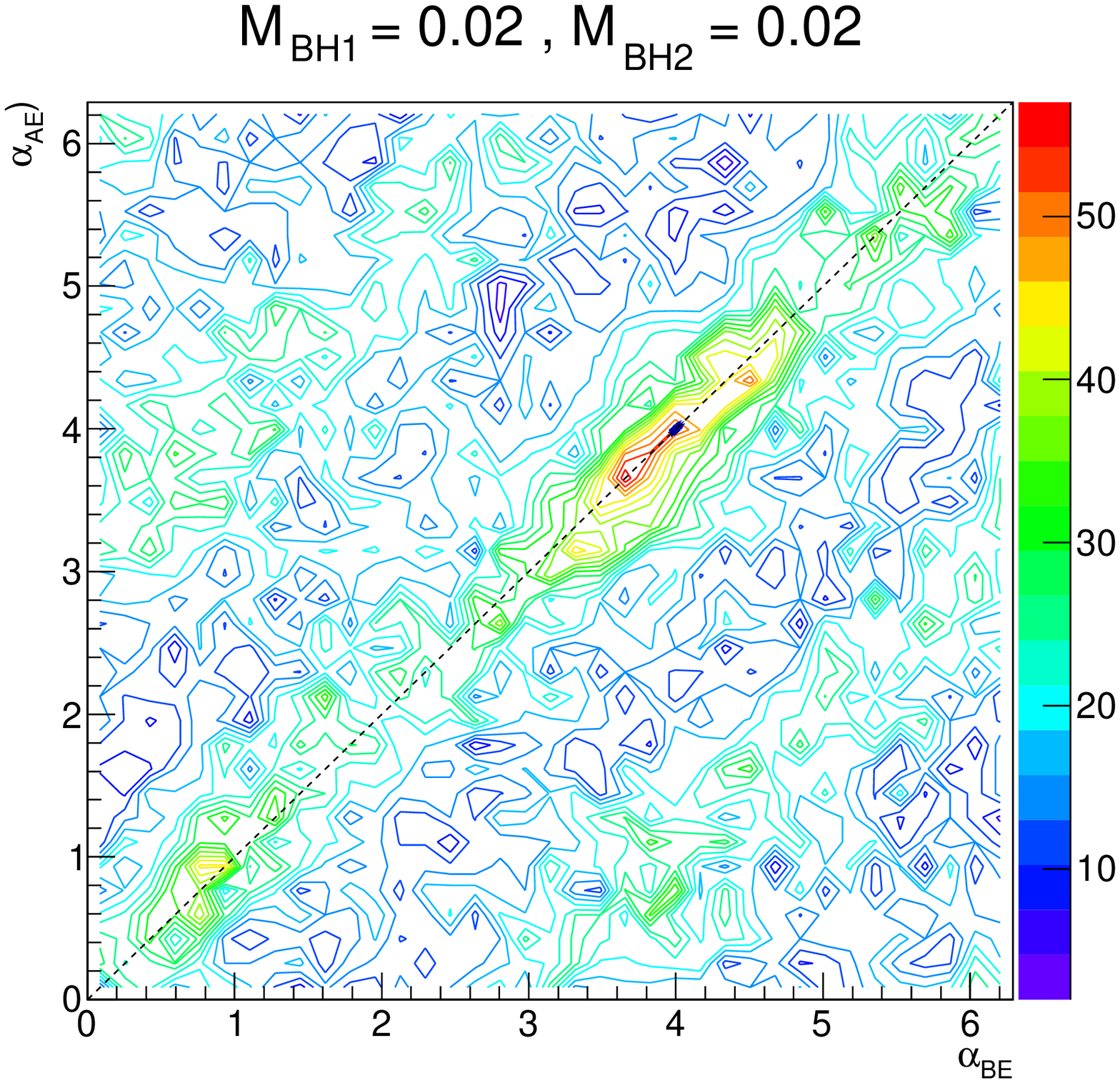}&
       \includegraphics[angle=0,width=1.0\columnwidth]{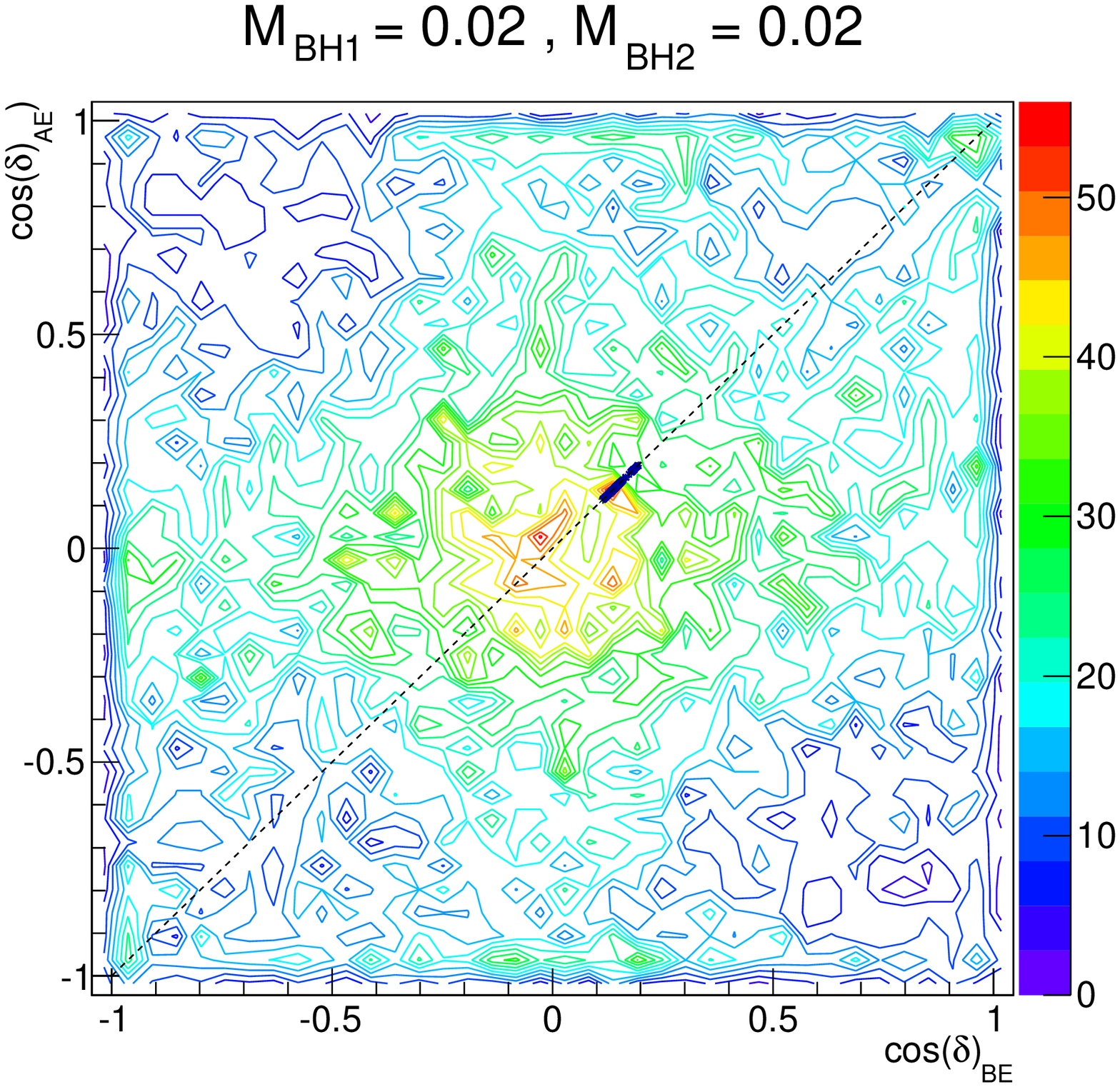}\\
       \includegraphics[angle=0,width=1.0\columnwidth]{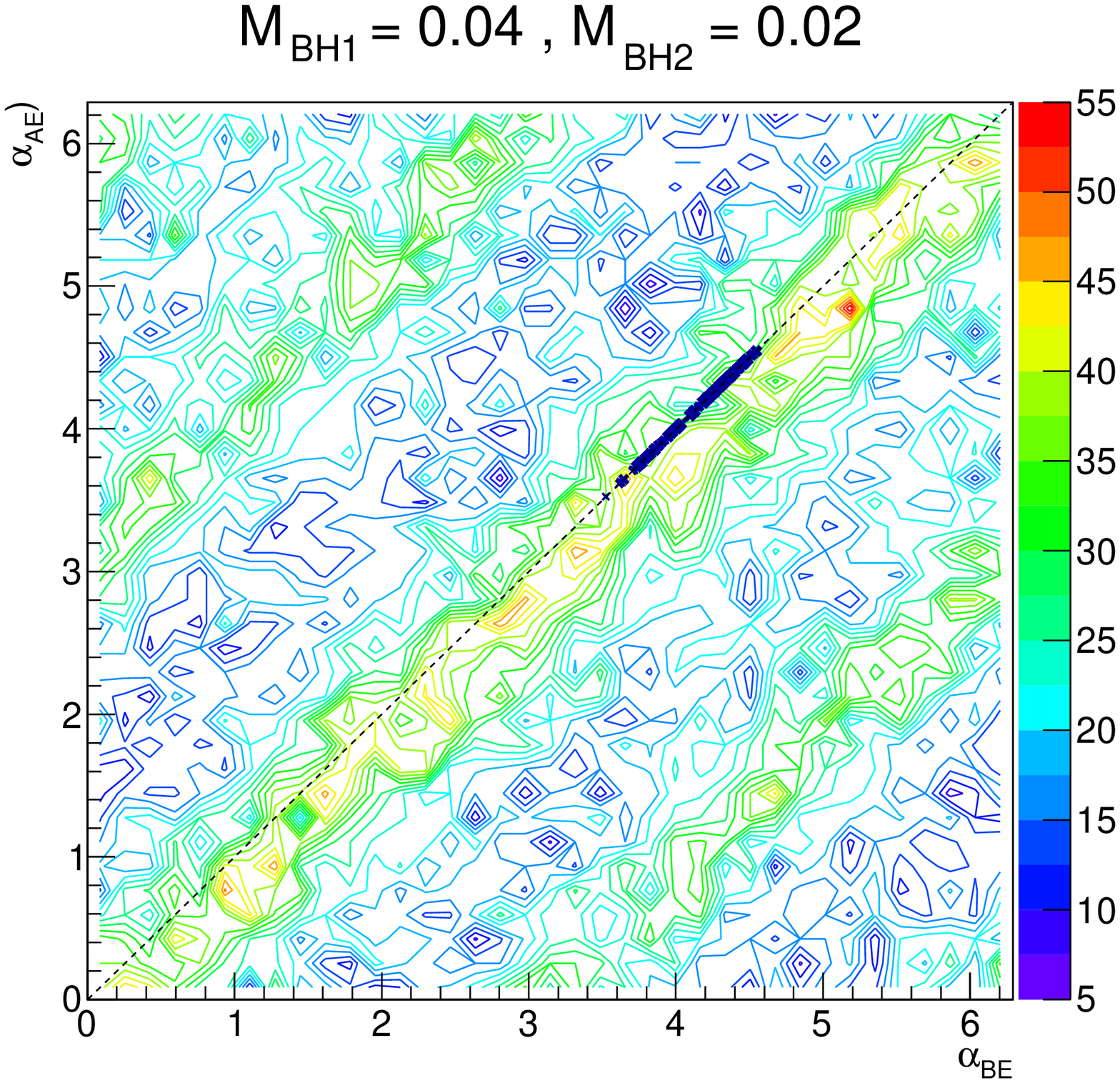}&
       \includegraphics[angle=0,width=1.0\columnwidth]{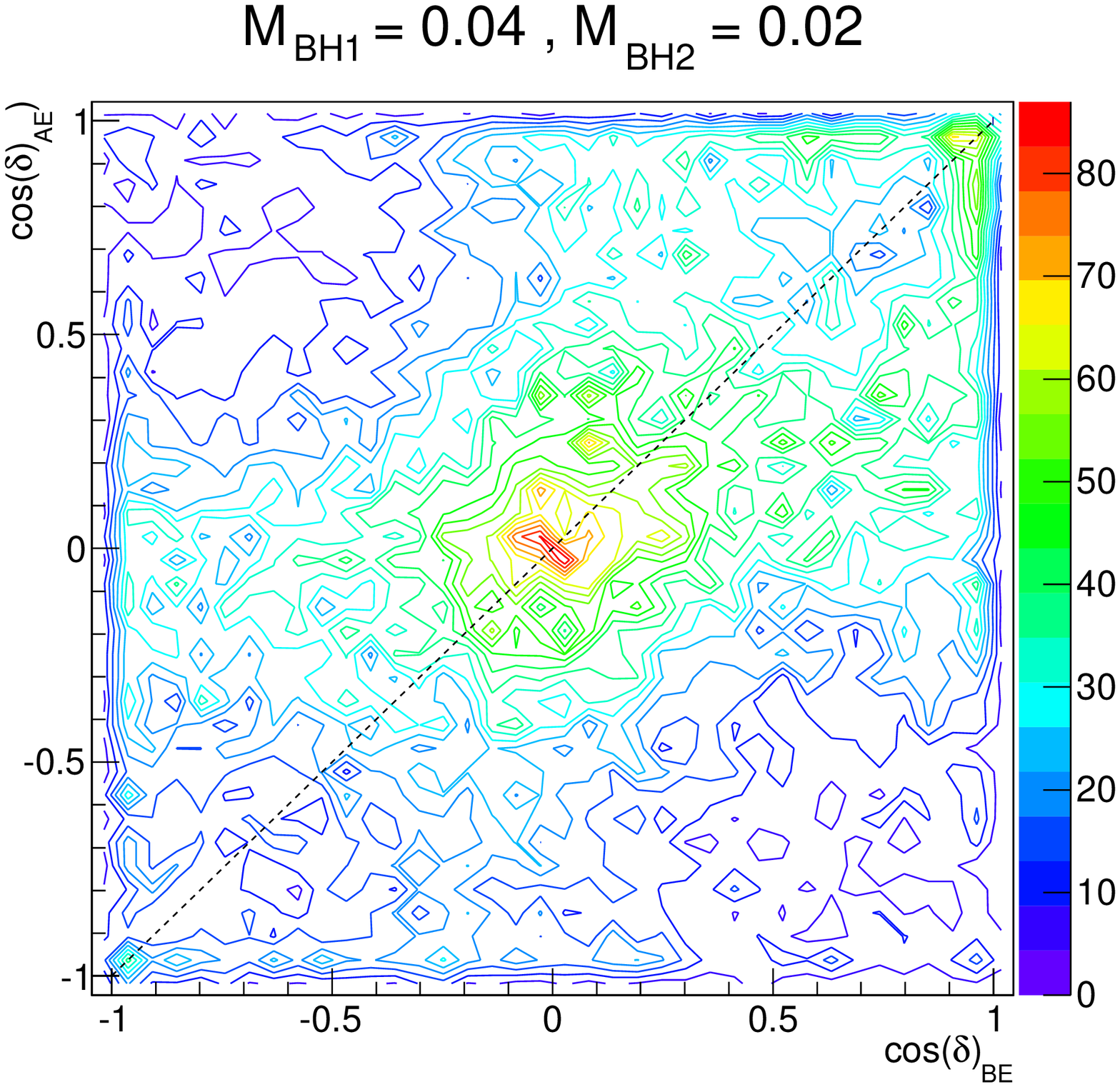}\\
     \end{tabular}
   \end{center}
   \caption{Left: the density map of $\alpha_{AE}$ vs. $\alpha_{BE}$ of ejected
     stars and MBHBs. Right: the similar map for $\delta$. Here are the same
     models $2020$ and $4020$ in Figure~\ref{fig:ad}. The colored contour indicate
     ejected stars and the black contour show MBHBs.}
   \label{fig:ad_io}
 \end{figure*}

 \begin{figure*}[htbp]
   \begin{center}
     \begin{tabular}{c c}
       \includegraphics[angle=0,width=1.0\columnwidth]{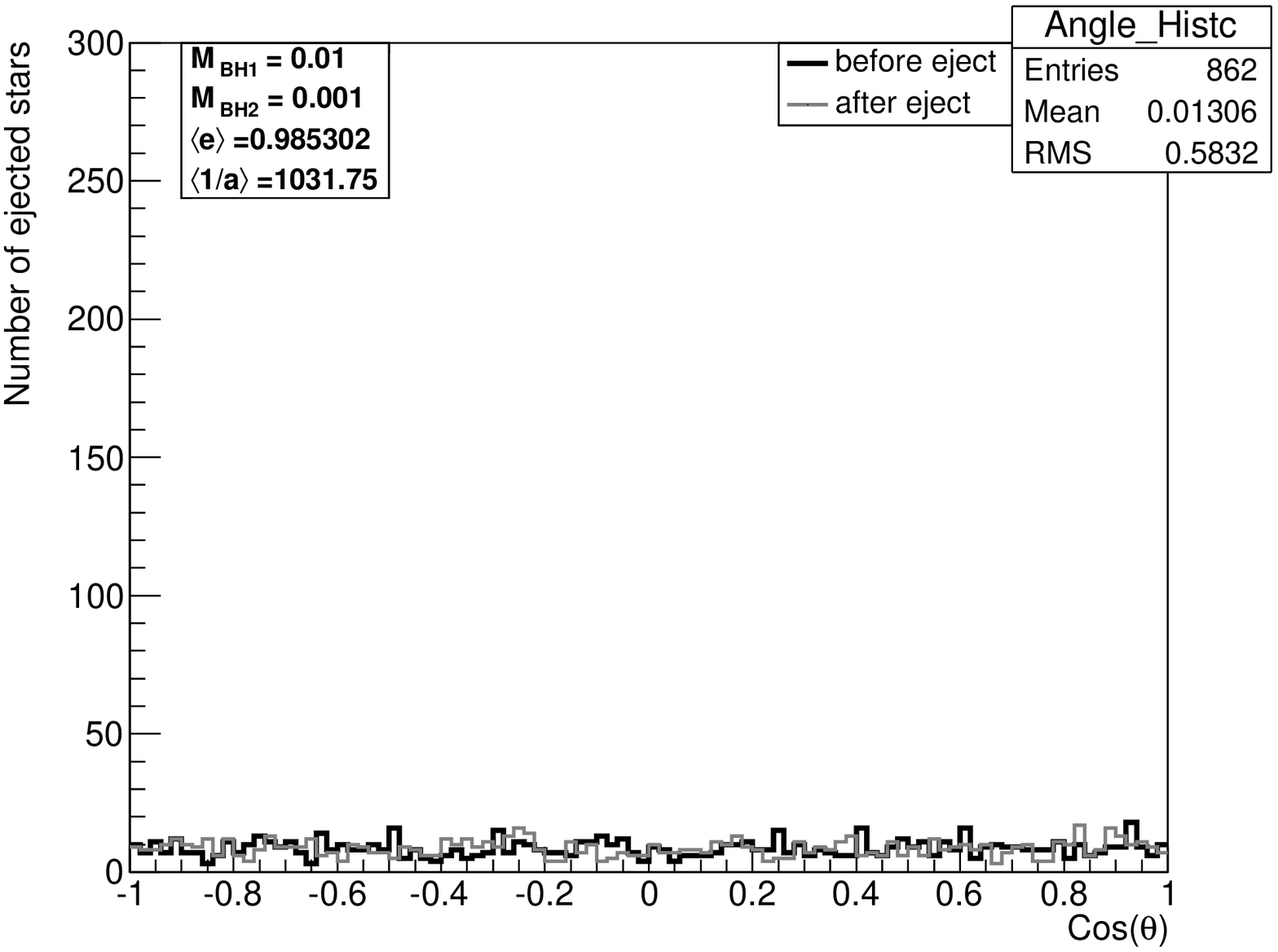}&
       \includegraphics[angle=0,width=1.0\columnwidth]{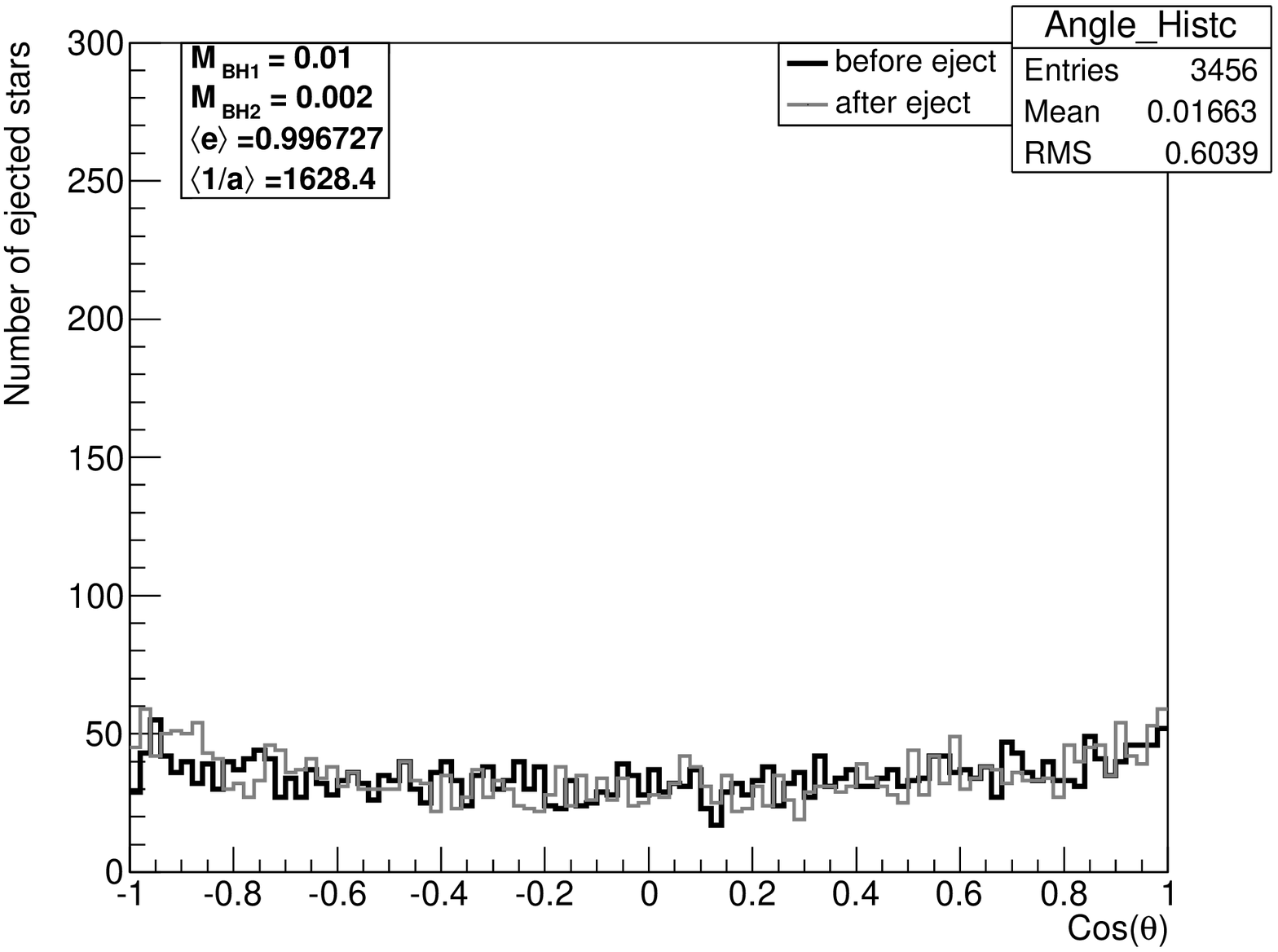}\\
       \includegraphics[angle=0,width=1.0\columnwidth]{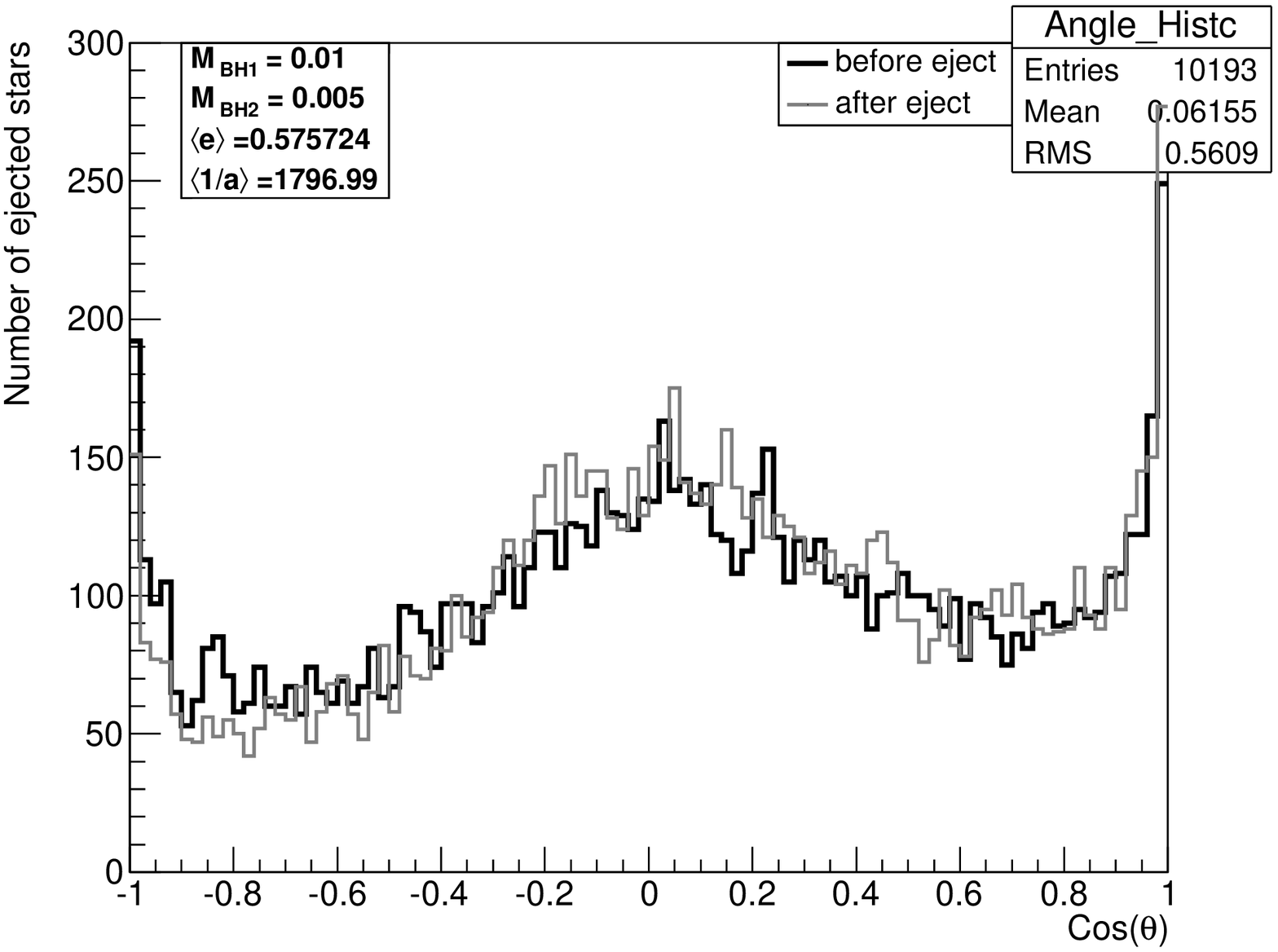}&
       \includegraphics[angle=0,width=1.0\columnwidth]{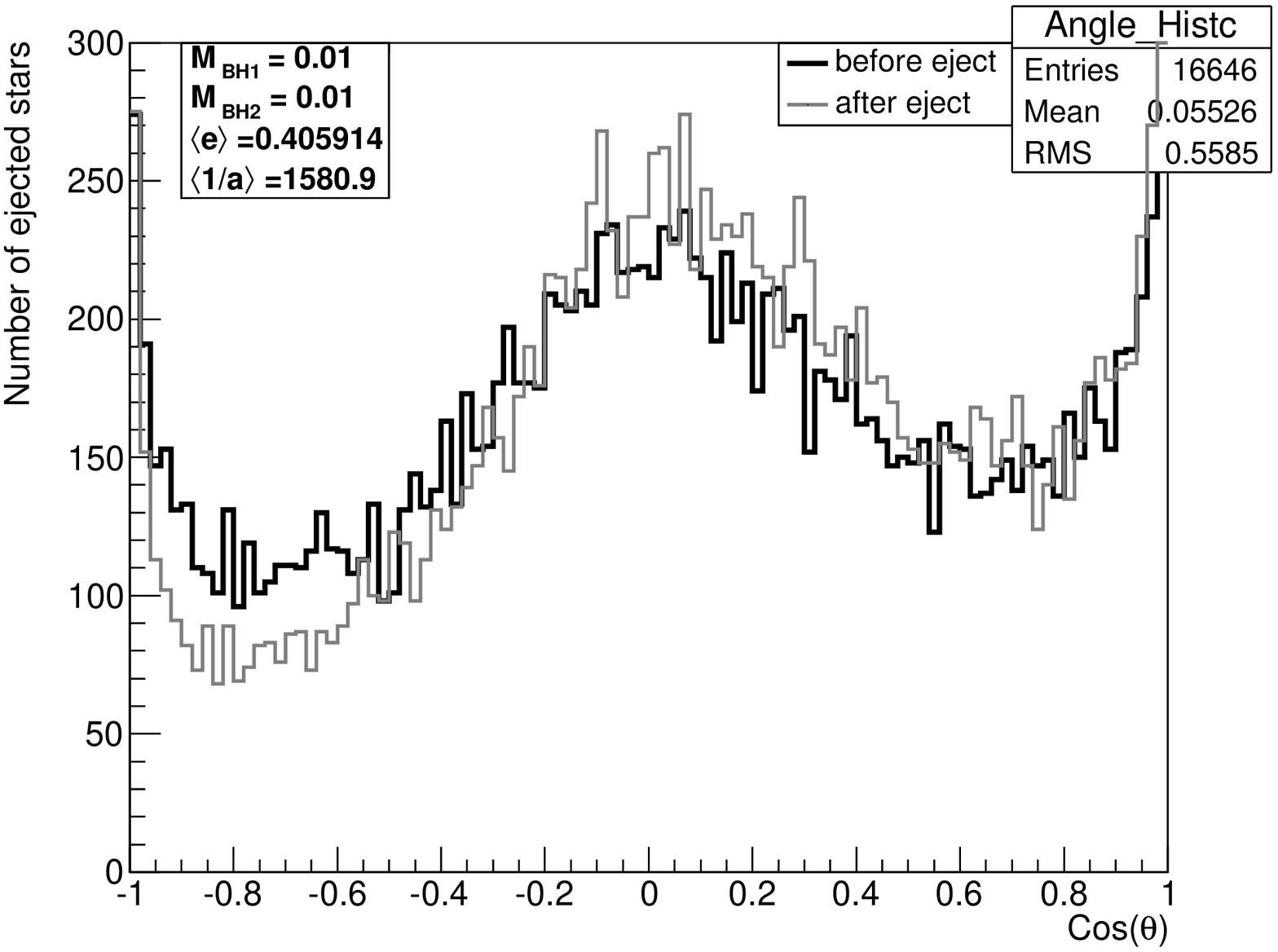}\\
     \end{tabular}
   \end{center}
   \caption{The distribution of $\cos (\theta)$ before and after ejection. In
     each sub-diagram, we also show the average eccentricity $\left\langle e
     \right\rangle$ and $\left\langle 1/a \right\rangle$ of MBHBs during time
     $100$ to $150$ in $N$-body units, where $a$ is semi-major axis of
     MBHBs. The last two panels are non-rotation models for comparison.}
   \label{fig:theta_io}
 \end{figure*}
\addtocounter{figure}{-1}
 \begin{figure*}[htbp]
   \begin{center}
     \begin{tabular}{c c}
       \includegraphics[angle=0,width=1.0\columnwidth]{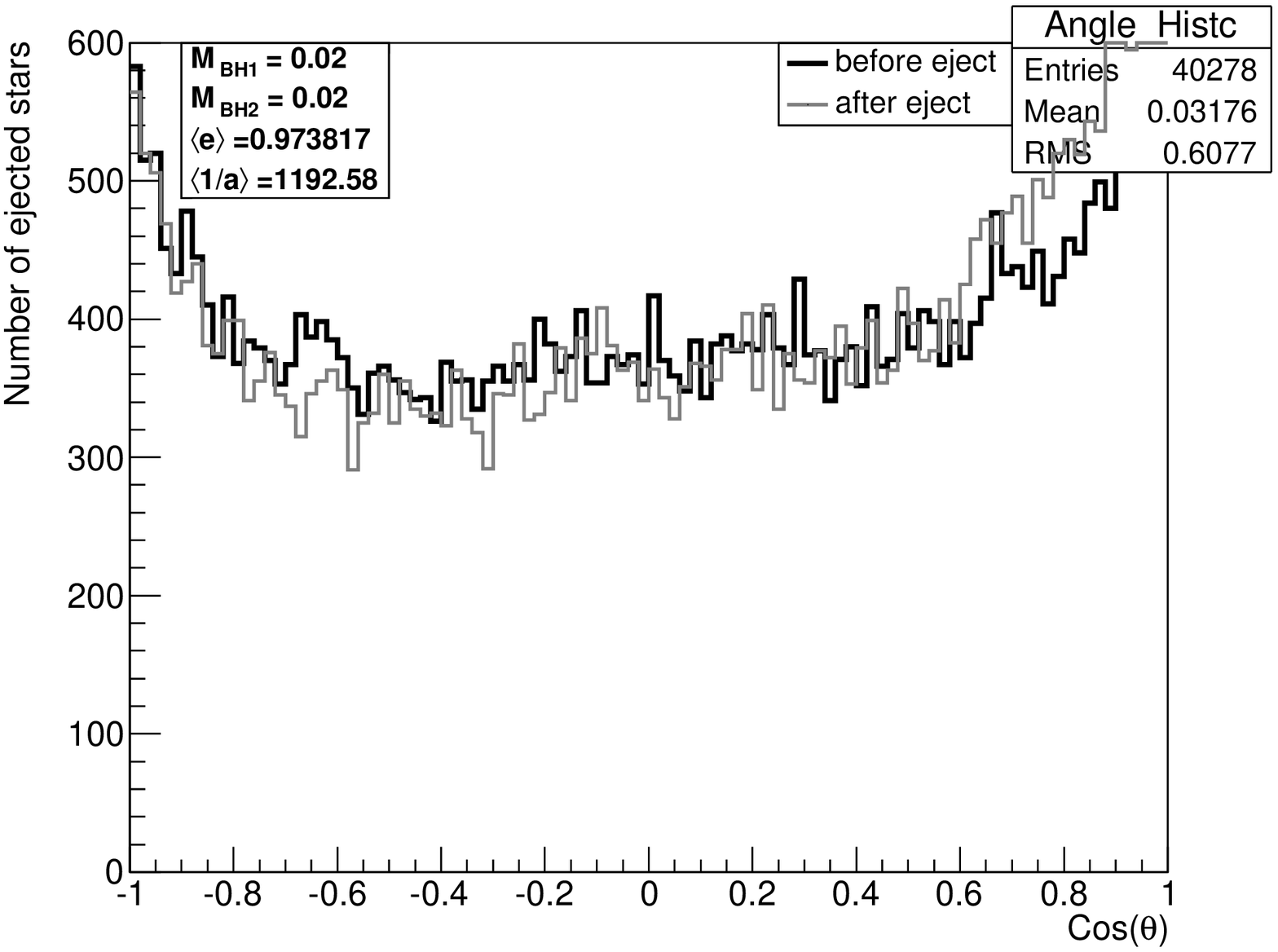}&
       \includegraphics[angle=0,width=1.0\columnwidth]{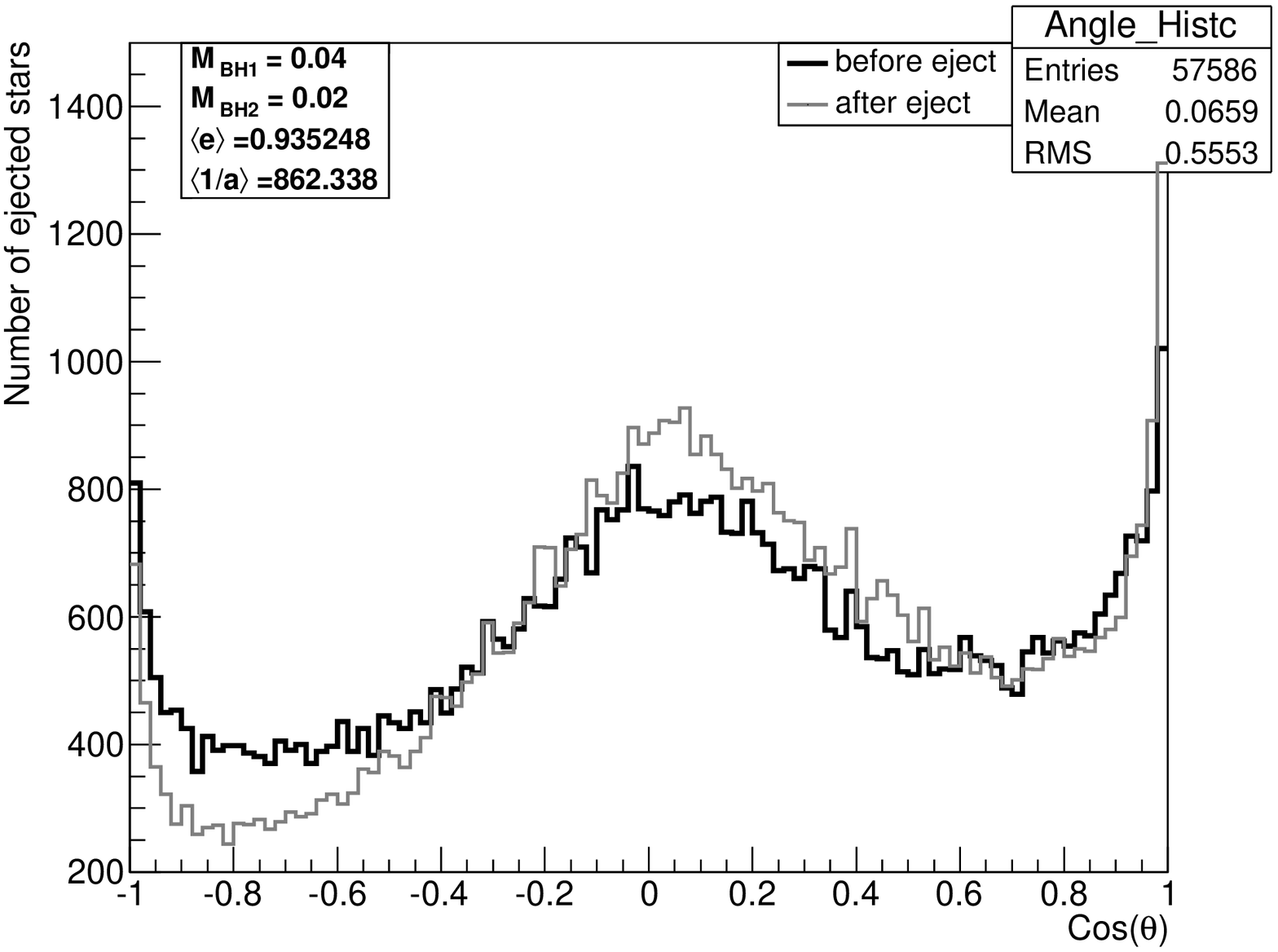}\\
       \includegraphics[angle=0,width=1.0\columnwidth]{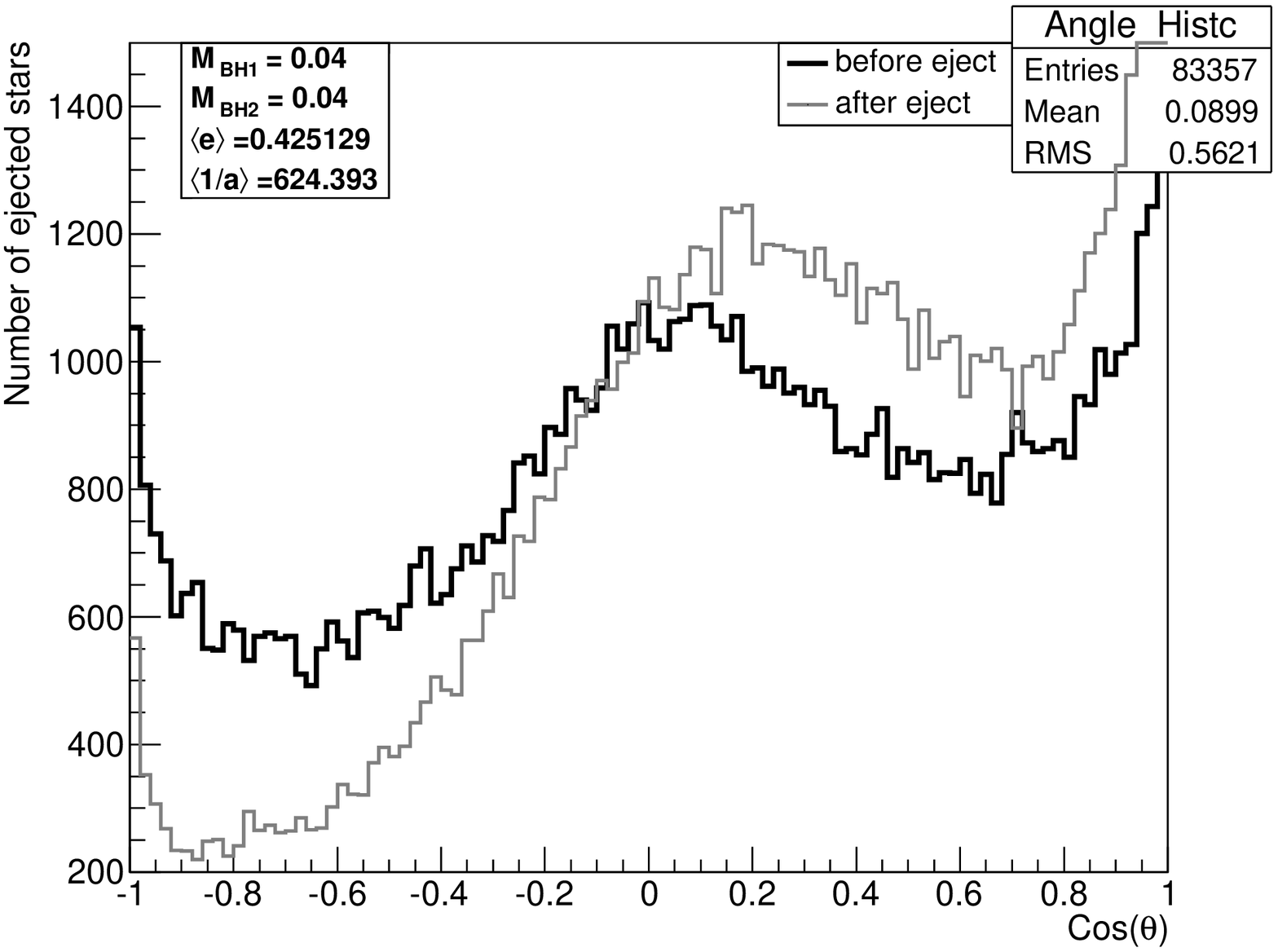}&\\
       \includegraphics[angle=0,width=1.0\columnwidth]{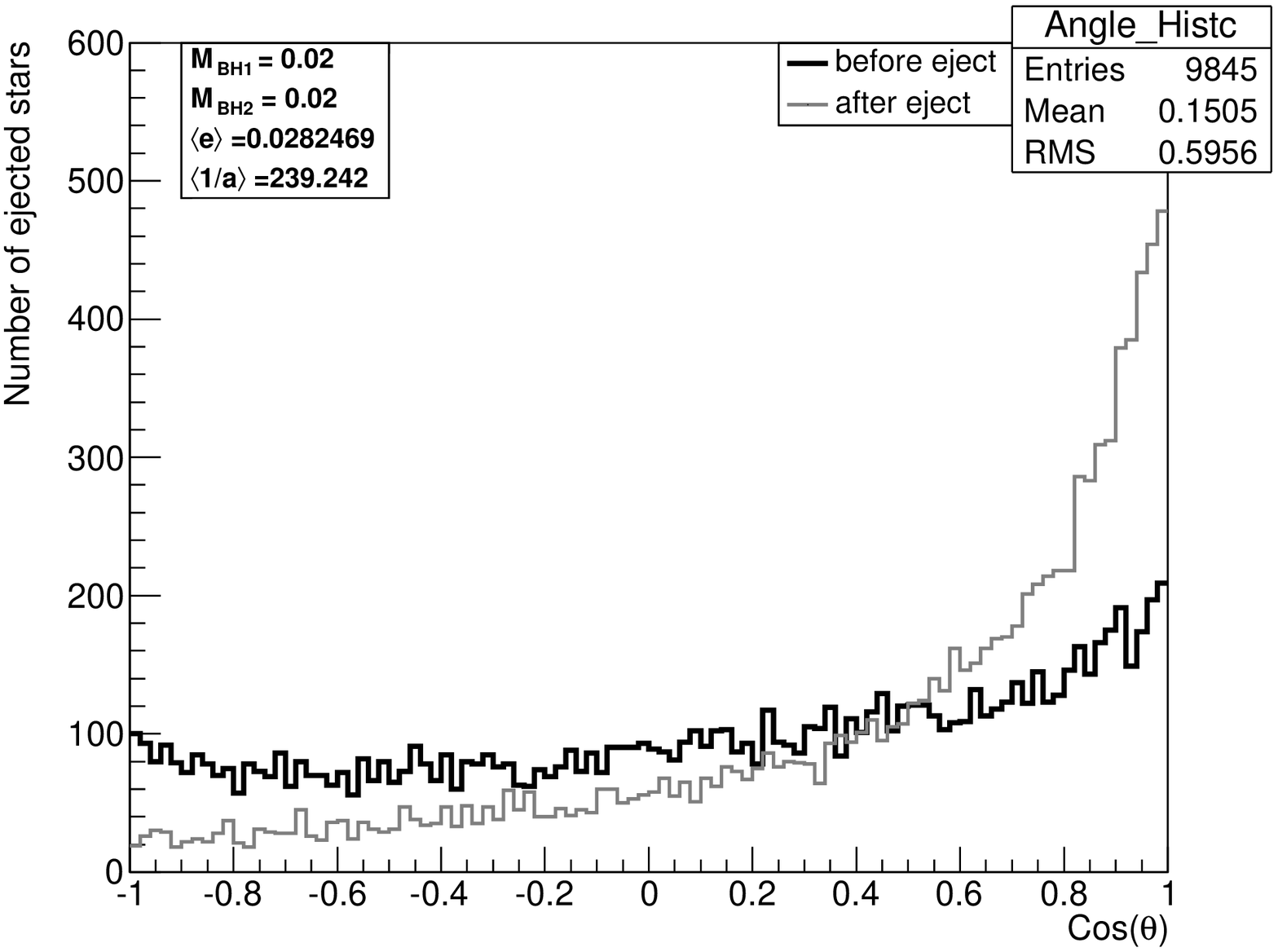}&
       \includegraphics[angle=0,width=1.0\columnwidth]{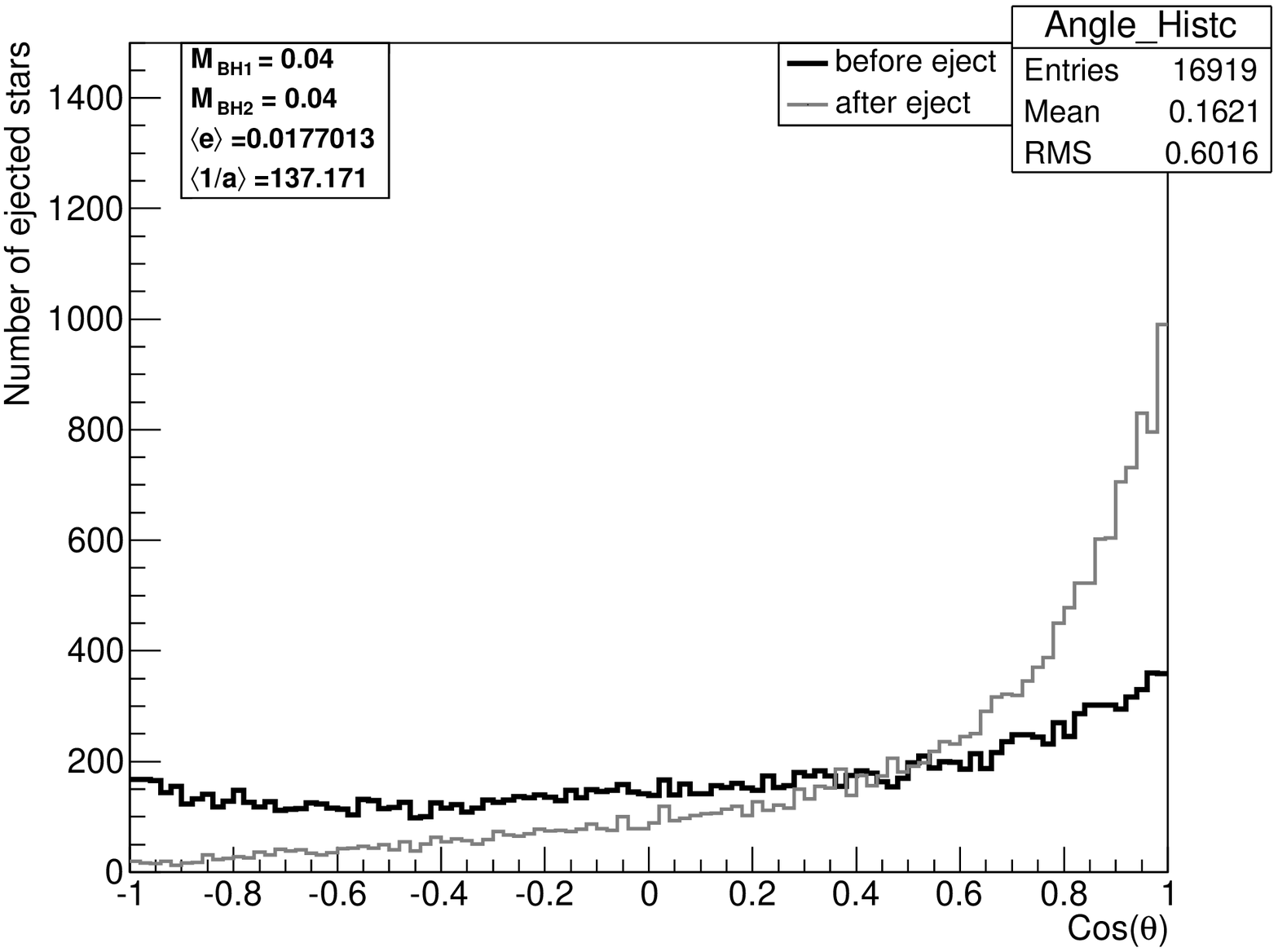}\\
     \end{tabular}
   \end{center}
   \caption{Continue of Figure~\ref{fig:theta_io}}
   \label{fig:theta_io:2}
 \end{figure*}

\end{document}